\documentclass[twocolumn]{article}
\pdfoutput=1
\usepackage[margin=0.8in]{geometry}
\usepackage[numbers, sort&compress]{natbib}
\usepackage{graphicx}
\usepackage{url}
\usepackage{amsmath}
\usepackage{color}
\usepackage{authblk}

\makeatletter
\renewcommand{\maketitle}{\bgroup\setlength{\parindent}{0pt}
  \textbf{\LARGE \@title}
\begin{flushleft}
  \@author
\end{flushleft}\egroup
\hfill \break
\hfill \break
\hfill \break
\hfill \break
\hfill \break
}
\makeatother

\title{The hydrophobic effect characterises the thermodynamic signature of amyloid fibril growth}

\author[1]{Erik van Dijk*}
\author[1]{Juami Hermine Mariama van Gils*}
\author[2]{Alessia Peduzzo}
\author[3]{Alexander Hofmann}
\author[3]{Georg Groth}
\author[4]{Halima Mouhib}
\author[5]{Daniel E. Otzen}
\author[2,6]{Alexander K. Buell}
\author[1]{Sanne Abeln}

\affil[1]{Computer Science Department, Center for Integrative Bioinformatics (IBIVU), VU University, Amsterdam, The Netherlands}
\affil[2]{Institute of Physical Biology, University of D\"usseldorf, Universit\"atsstr.1, 40225 D\"usseldorf, Germany}
\affil[3]{Institute of Biochemical Plant Physiology, University of D\"usseldorf, Universit\"atsstr.1, 40225 D\"usseldorf, Germany}
\affil[4]{Universit\'{e}e Paris-Est, Laboratoire Mod\'{e}lisation et Simulation Multi Echelle, MSME UMR 8208 CNRS, 5 bd Descartes, 77454 Marne-la-Vall\'{e}e, France}
\affil[5]{Interdisciplinary Nanoscience Center (iNANO) and Department of Molecular Biology, Gustav Wieds Vej 14, 8000 Aarhus C, Denmark}
\affil[6]{Department of Biotechnology and Biomedicine, Technical University of Denmark, S\o ltofts Plads, 2800 Lyngby, Denmark}

\begin{document}

\begin{titlepage}
\maketitle

%\onecolumn
%\begin{itemize}
% \item[$^1$] Dep. of Computer Science, Vrije Universiteit Amsterdam, The Netherlands
% \item[$^2$] Institute of Physical Biology, University of D\"usseldorf, Universit\"atsstr.1, 40225 D\"usseldorf, Germany
% \item[$^3$] Institute of Biochemical Plant Physiology, University of D\"usseldorf, Universit\"atsstr.1, 40225 D\"usseldorf, Germany
% \item[$^4$] Universit\'{e}e Paris-Est, Laboratoire Mod\'{e}lisation et Simulation Multi Echelle, MSME UMR 8208 CNRS, 5 bd Descartes, 77454 Marne-la-Vall\'{e}e, France
% \item[$^5$] Interdisciplinary Nanoscience Center (iNANO) and Department of Molecular Biology, Gustav Wieds Vej 14, 8000 Aarhus C, Denmark
% \item[$^6$] Current address: Department of Biotechnology and Biomedicine, Technical University of Denmark, S\o ltofts Plads, 2800 Lyngby, Denmark
%\end{itemize}

% Abstract
%\include{abstract_long}
%\include{abstract_short}

\begin{abstract}
Many proteins have the potential to aggregate into amyloid fibrils, which are associated with a wide range of human disorders including Alzheimer’s and Parkinson’s disease. In contrast to that of folded proteins, the thermodynamic stability of amyloid fibrils is not well understood: specifically the balance between entropic and enthalpic terms, including the chain entropy and the hydrophobic effect, are poorly characterised. Using simulations of a coarse-grained protein model we delineate the enthalpic and entropic contributions dominating amyloid fibril elongation, predicting a characteristic temperature-dependent enthalpic signature. We confirm this thermodynamic signature by performing calorimetric experiments and a meta-analysis over published data. From these results, we can also elucidate the necessary conditions to observe cold denaturation of amyloid fibrils. Overall, we show that amyloid fibril elongation is associated with a negative heat capacity, the magnitude of which correlates closely with the hydrophobic surface area that is buried upon fibril formation, highlighting the importance of hydrophobicity for fibril stability.
\end{abstract}

%Many proteins have the potential to aggregate into amyloid fibrils, which are associated with a wide range of human disorders including Alzheimer’s and Parkinson’s disease. In contrast to that of folded proteins, the thermodynamic stability of amyloid fibrils is not well understood: specifically the balance between entropic and enthalpic terms, including the chain entropy and the hydrophobic effect, are poorly characterised. Here we use simulations that can capture these terms explicitly in terms of their enthalpic and entropic contributions. We also perform calorimetric experiments at different temperatures and gather available data from the literature to investigate the contribution of the thermodynamic components. From our results, we elucidate the necessary conditions to observe cold denaturation of amyloid fibrils. We show that amyloid fibril elongation is generally associated with a large and negative heat capacity, the magnitude of which correlates closely with the hydrophobic surface area that is buried upon fibril formation.

\end{titlepage}
%\newpage

\section*{Introduction}
\begin{figure}[t]
  \includegraphics[width=0.95\linewidth]{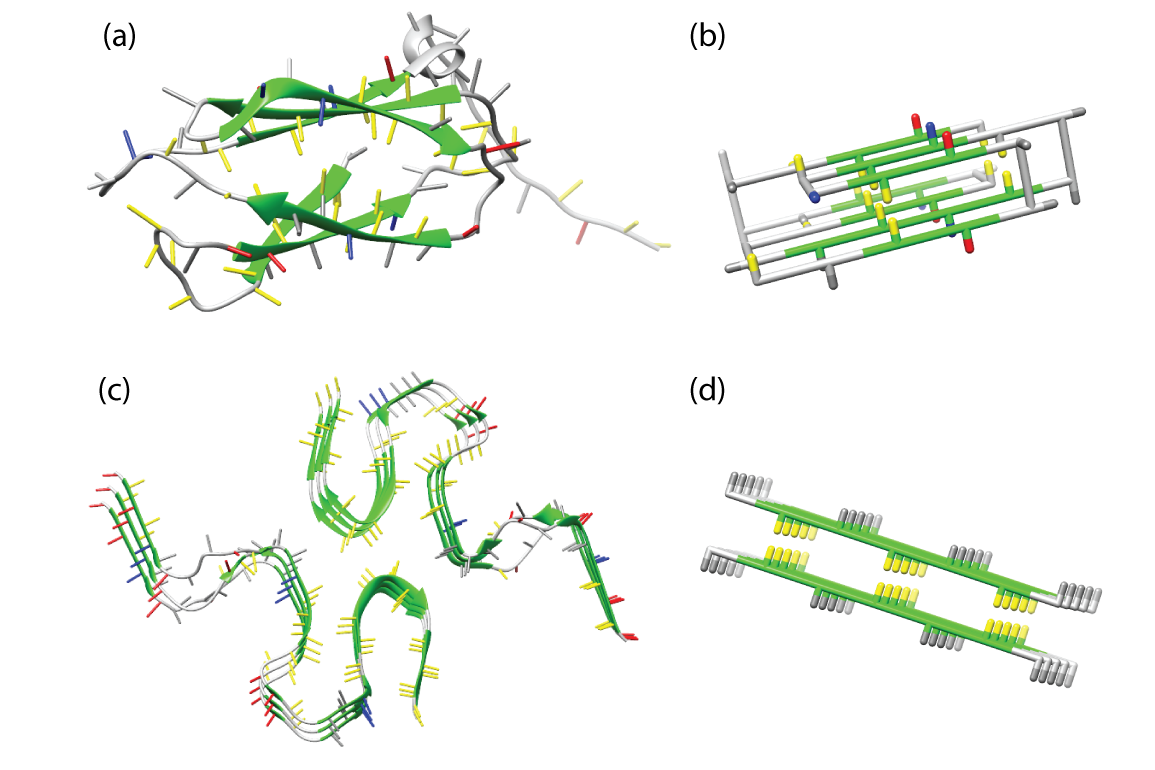}
  \caption{\textbf{Folded and fibrillar states}. (a) The native state of human transglutaminase (PDB-ID: 2XZZ) (b) a model protein in the native state, (c) disease-related amyloid $\beta$-sheet of the A$\beta$ (1-42) peptide (PDB-ID: 2NAO)~\cite{Walti2016} and (d) the seed structure of an amyloid fibril represented in the lattice model used for this work. Side chains of hydrophobic residues are coloured in yellow, of polar residues in grey, of positively charged residues in blue and of negatively charged in red. For $\beta$-stranded structures the backbone is coloured in green. In the folded protein, a hydrophobic core can be observed, where the hydrophobic residues are shielded from the water by the hydrophilic and charged residues. Similarly, the core sequence regions of amyloid fibril structures tend to be strongly hydrophobic.
  }
  \label{fig:intro}
\end{figure}

The folding of proteins is an essential process for cellular functioning. Protein misfolding and aggregation on the other hand can severely deregulate cells. The most dominant contributions to protein folding include pairwise amino acid interactions, the hydrophobic effect and the configurational entropy; these contributions can be captured by a range of models describing the process at different temperatures: \cite{Shakhnovich1993a, %Shakhnovich1993b, 
Morris-Andrews2015, %Shakhnovich1994, 
Sali1994, Coluzza2003, Coluzza2004, Coluzza2007, 
Abeln2011, Vacha2014, vanDijk2016, Eugene2016, Saric2016, Smit2017, Dijkstra2018}. 
The protein folding process involves several enthalpy-entropy compensating mechanisms. The hydrophobic effect, the main stabilizing factor in the folded state ~\cite{Baldwin2007},  contains both enthalpic and entropic components \cite{Widom2003, Chandler2005, Liu2005}; this is apparent from the weakening of this effect at low temperatures \cite{Dias2010}. Similarly, the chain configurational entropy and the pairwise amino acids interaction can lead to enthalpy-entropy compensation; this is apparent from the heat-induced unfolding of proteins \cite{Shakhnovich1993a, %Shakhnovich1993b, 
Morris-Andrews2015, %Shakhnovich1994, 
Sali1994, Coluzza2003, Coluzza2004, Coluzza2007}.
The accurate description of the hydrophobic effect within several models for protein folding, using both the enthalpic and entropic terms, has enabled the reproduction of experimentally observed features, such as cold-, heat-, and pressure-induced denaturation~\cite{Chan1998, Kaya2003, Huang2000, Wolde2002, Sirovetz2015, vanDijk2015, vanDijk2016, Brotzakis2016, Bianco2017a, Pucci2017}, thereby emphasizing also the role of hydrophobicity in the stability of folded proteins (see Figure \ref{fig:intro} (a) for an example of a folded protein with a hydrophobic core). Thus, the enthalpy-entropy compensation mechanisms in protein folding are well-understood and important for rationalising the thermodynamic characteristics of protein folding. 

Under certain conditions, specific proteins can aggregate into $\beta$-strand-dominated amyloid fibrils. This process is believed to be the underlying cause of many degenerative diseases, such as the A$\beta$ peptide in the case of Alzheimer's disease and $\alpha$-synuclein in the case of Parkinson's disease \cite{Varadi2018}. Unlike protein folding, the fibril formation process is not well-understood in terms of its thermodynamic characteristics.

Recently resolved full length fibril structures  \cite[e.g.][]{Walti2016, Tuttle2016} suggest that the cores of disease-associated amyloid fibril structures are characterised by a higher average hydrophobicity than the overall sequences, see Figure \ref{fig:intro} (c), indicating that hydrophobicity may play an important role in the formation of amyloid fibrils.
We can observe thermodynamic properties of amyloid fibrils experimentally at different temperatures, for example by using Differential Scanning Calorimetry (DSC,~\cite{Sasahara2005, Morel2010}) or Isothermal Titration Calorimetry (ITC)~\cite{Kardos2004, Jeppesen2010, Ikenoue2014}. Under constant pressure, the heat ($dQ$) added or removed from the system is equal to its change in enthalpy ($dH$). ITC experiments show that amyloid fibril growth is generally an exothermic process~\cite{Kardos2004, Jeppesen2010} with a negative heat capacity \cite{Jeppesen2010, Ikenoue2014}. However, a few notable exceptions with positive heat capacity have been reported \cite{Jeppesen2010, Ikenoue2014}.
Under physiological conditions, amyloid fibrils are often very stable \cite{Baldwin2011}, but high free energy barriers prevent aggregates from forming~\cite{Buell2012}. In addition to heat denaturation \cite{Sasahara2005, Morel2010}, some fibrils also denature at temperatures near the freezing point~\cite{Kim2008, Ikenoue2014}. 
In contrast to protein folding, it is unclear how the different enthalpic and entropic contributions may lead to these temperature dependent effects, and how thermodynamic signatures of fibril elongation may be rationalised. In order to elucidate the molecular origins of the stability of amyloid fibrils and the associated thermodynamic signatures, insight from computational models is required.

Here, we use Monte Carlo simulations of a coarse-grained physical protein model situated on a cubic lattice to study amyloid fibril elongation. %established for the study of protein aggregation~\cite{Abeln2014, Ni2013, Ni2015, Tran2016, Chiricotto2017} that we combine with a method to account for the temperature dependence of the hydrophobic effect \cite{Abeln2011, vanDijk2016}. 
Previously, we included the temperature dependence of the hydrophobic effect in a lattice model to delineate the enthalpic and entropic contributions of protein folding \cite{Abeln2011,vanDijk2016}; however, this classic lattice model is unsuitable to study amyloid fibril formation.  We also developed a different model that includes hydrogen-bond dependent beta-strand formation \cite{Abeln2014} to simulate amyloid fibril formation \cite{Abeln2014, Ni2013, Ni2015, Tran2016, Chiricotto2017}. The coarse-grained lattice model used in this work therefore incorporates elements of these two previous models: hydrogen-bond dependent beta-strand formation \cite{Abeln2014} and the temperature dependence of the hydrophobic effect \cite{vanDijk2016}. It explicitly captures the dominating enthalpic and entropic components, including: chain entropy, entropy-enthalpy compensation from the hydrophobic effect, and enthalpic terms from hydrogen bonds and side-chain interactions. Figure \ref{fig:intro} (b) shows a folded protein represented by the model.

In this study, we focus on the thermodynamic signatures and their temperature dependence associated with the elongation of a preformed amyloid fibril, as shown in Figure \ref{fig:intro} (d). We do not consider the kinetic or thermodynamic characteristics of the nucleation processes and the possible formation of oligomers as intermediate states before the formation of well-defined fibrils. The focus on the elongation step allows us to simplify the simulation setup, and to compare our simulations directly with data from microcalorimetry and equilibrium experiments from the literature and from our own experiments. We have gathered an extensive collection of thermodynamic data of peptide assembly and we find that the predictions of our model are corroborated by the available data. 
Using this approach, we show that cold denaturation only occurs when (1) the hydrophobic effect represents a significant component of the fibril stability, (2) the overall stability of the fibril is sufficiently low and (3) fibril elongation changes from being an endothermic reaction at lower temperatures to an exothermic reaction at higher temperatures. Furthermore, we find that the strength of the temperature effect on the enthalpy of fibril elongation (i.e. the absolute value of the heat capacity) is directly dependent on the hydrophobicity of the aggregating region of the fibril.
Finally, we show that in addition to the hydrophobic effect, the backbone entropy has a significant influence on amyloid fibril stability.

\section*{Results and Discussion}
In this work, we aim to investigate the effect of interplay between the hydrophobic temperature dependence, configurational chain entropy and pairwise enthalpic interactions between amino acids (including hydrogen bonds) on the thermodynamic characteristics of amyloid fibril elongation.
First, we consider if the addition of the hydrophobic effect could lead to a destabilisation of an amyloid fibril at low temperatures and ultimately to cold denaturation. 

\begin{figure*}[t]
 \includegraphics[width=\linewidth]{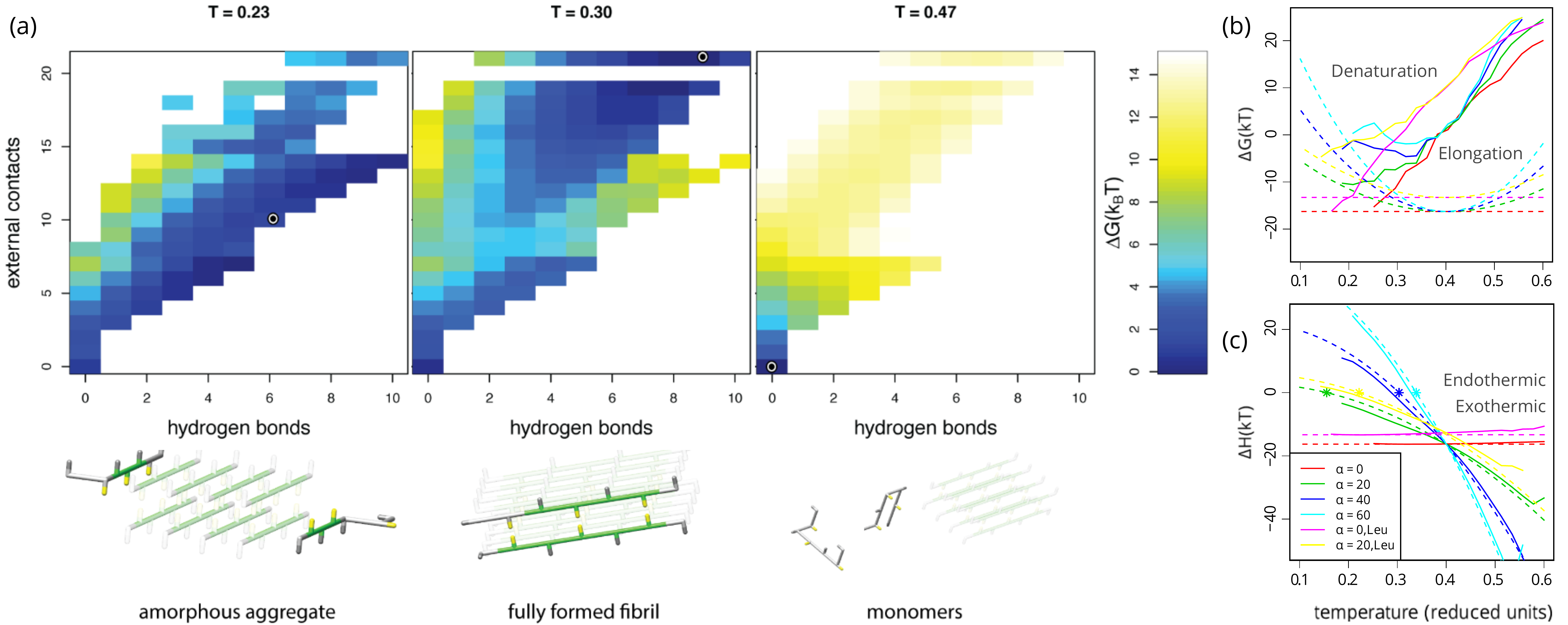}
  \caption{{\bf Temperature dependent states} (a) Top: free energy landscapes, and bottom: representative snapshots are shown for simulations at three different temperatures modelled with a strong hydrophobic temperature dependence, $\alpha$ = 60. The snapshots of the peptides are representative for the low free energy conformations; the circle represents the exact order parameters for the snapshot in the corresponding free energy landscape. Within the snapshots the seed fibril is indicated in a lighter shade (see caption Figure \ref{fig:intro} for colour coding). 
  At high temperatures only transient (external) contacts are formed and the protein molecules that are not part of the seed remain in their monomeric form. At intermediate temperatures the free peptides adopt a regular, fibrillar structure at the end of the seed. At low temperatures the heatmap shows that there is not a single distinct conformation with a low free energy: the simulated free peptides are attached to the seed fibril, but no hydrophobic core is formed between the peptides. 
  (b) Free energy and (c) Enthalpy of fibril elongation. The differences between the fully formed fibril ($C_\text{ext}$=21) and the monomeric state ($C_\text{ext}$=0) are calculated from the simulation of several different fibrils. The interaction strength of the Leu based fibrils is weaker, leading to a lower enthalpic contribution, effectively weakening the fibrils. Dotted lines indicate estimates for the hydrophobic contributions showing from left to right $\Delta \hat{G}_{\text{hydr}}$ and $\Delta \hat{E}_{\text{hydr}}$; these estimates are generated using Eqns. \ref{eq:dG} and \ref{eq:dE} with corresponding $\alpha$, $\Delta C_h=-6$ and with an offset, $E_\text{int}=\Delta H$ based on simulations with the equivalent peptide for $\alpha=0$; stars indicate the change of an exothermic to an endothermic process, based on the $\Delta \hat{E}_{\text{hydr}}$ estimate. 
  It is clear that in our model the slope of the enthalpy of fibril elongation as a function of temperature, is dominated by the temperature dependence of the hydrophobic effect.
  }
  \label{fig:heatmap}
\end{figure*}

Here, we use a simple model of a short (seed) fibril that is made up of two layers of peptides with alternating hydrophobic and hydrophilic amino acids~\cite{Abeln2014}. The fibril structure has a hydrophobic core between the two layers, as shown in Figure \ref{fig:intro} (d). We sample this model by considering the process of fibril elongation: two peptides of the fibril are free to restructure in the simulations, while the remaining peptides in the seed fibril remain structurally fixed (Figure \ref{fig:heatmap} (a)). Finally, we add an explicit temperature dependence for hydrophobicity \cite{vanDijk2016}. In  simulations of our model of peptide aggregation, all fibrils denature into monomers at high temperatures, as shown in Figures \ref{fig:heatmap} and \ref{fig:characteristicsPlot}, consistent with previously our expectations of this model~\cite{Abeln2014}. Experimentally, heat denaturation has also been shown for several types of fibrils~\cite{Sasahara2005, Ikenoue2014} and is even used for the destruction of contaminating amyloid fibrils, such as prions, in a clinical context~\cite{McDonnell2003}.

\subsection*{Cold destabilisation and denaturation}
To investigate the effect of the hydrophobic temperature dependence, we vary a parameter $\alpha$ that sets the strength of the hydrophobic effect, see equation \ref{eq:hG} in the Methods; a similar parameter was used in earlier work to model the cold denaturation of folded proteins~\cite{vanDijk2016}. Destabilisation of the fibril at lower temperatures is only observed when a moderately strong temperature dependence of the hydrophobic effect is included in the model ($\alpha=40$ and $\alpha=60$). In our simulations with $\alpha = 60$, we observe heat denaturation, as well as cold-destabilisation and even cold denaturation of the fibril into a less ordered aggregated (`amorphous'), as well as into the monomeric state ($\alpha=60$, $T \approx 0.25$), see Figure \ref{fig:characteristicsPlot}. 
When we investigated the aggregated and denatured states in more detail (Figure \ref{fig:heatmap} (a)), we found that the denatured state at low temperatures is structurally distinct from the heat denatured state: the low temperature amorphous structural ensemble displays more residual interactions (left panel Figure \ref{fig:heatmap} (a)). This is consistent with experimental results showing that under some conditions, oligomeric states can be found at low temperatures~\cite{Kim2008}, as opposed to the monomeric protein molecules that are found at very high temperatures~\cite{Sasahara2005, Ikenoue2014}. The more compact configurational ensemble of the cold denatured state of a fibril shows clear parallels with cold denaturation of monomeric proteins \cite{Vajpai2013, vanDijk2016}. 

\subsection*{Thermodynamic Signature of Cold Denaturation}
To shed more light on the nature of the cold denaturation transition, we consider the free energy difference between the fibrillar and monomeric states. When the free energy of the monomeric, or denatured state becomes lower than the fibrillar state, the fibril will dissolve; this corresponds to a $\Delta G > 0$ in Figure \ref{fig:heatmap} (b). When we consider the enthalpic and entropic contributions of fibril stability separately, in Figure \ref{fig:heatmap} (c) and \ref{fig:TdS} respectively, it is clear that these individual contributions are large compared to the overall difference in free energy. Similar observations on enthalpy-entropy compensation have for example been made for both the stability of folded proteins~\cite{Liu2000}, and the free energy barriers of amyloid fibril growth~\cite{Buell2012}.

Moreover, comparing the different contributions to the free energy of elongation it is clear that the configurational chain and diffusional entropy that is lost upon binding dominates the free energy difference at high temperatures, causing the heat denaturation of the fibril. This can be most clearly seen comparing Figure \ref{fig:heatmap} (b) and Figure \ref{fig:TdS} for $\alpha=0$. 

For cold denaturation on the other hand, it is the hydrophobic temperature dependence that dominates this transition (cyan curves in Figure \ref{fig:heatmap} (b) and (c)). Note that free energy and entropy components still contain rather large configurational contributions at these low temperatures; they do not precisely follow the hydrophobic components.

Nevertheless, the enthalpy of fibril growth strongly correlates with the signature of the hydrophobic effect ($\Delta \hat{E}_{hydr}$), as seen in Figure \ref{fig:heatmap} (c). The temperature independent potential ($\alpha$ = 0) leads to an approximately constant enthalpy of fibril elongation. Adjusting the strength of the hydrophobic effect through a change of $\alpha$ leads to a negative slope in the enthalpy as a function of temperature, which corresponds to a negative heat capacity of the elongation reaction, since $\frac{\partial \Delta E_{el}}{\partial T}$=$\Delta$C$_{p,el}$. Hence, the hydrophobic temperature dependence can directly explain the negative slope of the enthalpy of fibril elongation versus temperature.

\subsection*{Comparison to Experimental Signatures}
In our model, the inclusion of the hydrophobic temperature dependence effectively yields a negative value for heat capacity of the elongation reaction, where we assume $\Delta$C$_{v,el} \approx \Delta$C$_{p,el}$ as in Ref.~\cite{vanDijk2016}. 

A negative value of $\Delta$C$_{p,el}$ is found for most amyloid fibrils in experimental work~\cite{Kardos2004, Jeppesen2010, Ikenoue2014}, consistent with an increasingly exothermic signature of fibril elongation as the temperature increases. Interestingly, however, it was reported that the fibril elongation of $\alpha$-synuclein~\cite{Ikenoue2014}, as well as glucagon at moderate to high ionic strength~\cite{Jeppesen2010} has a small positive heat capacity.

\begin{figure*}[t]
  \centering
  \includegraphics[width=.95\linewidth]{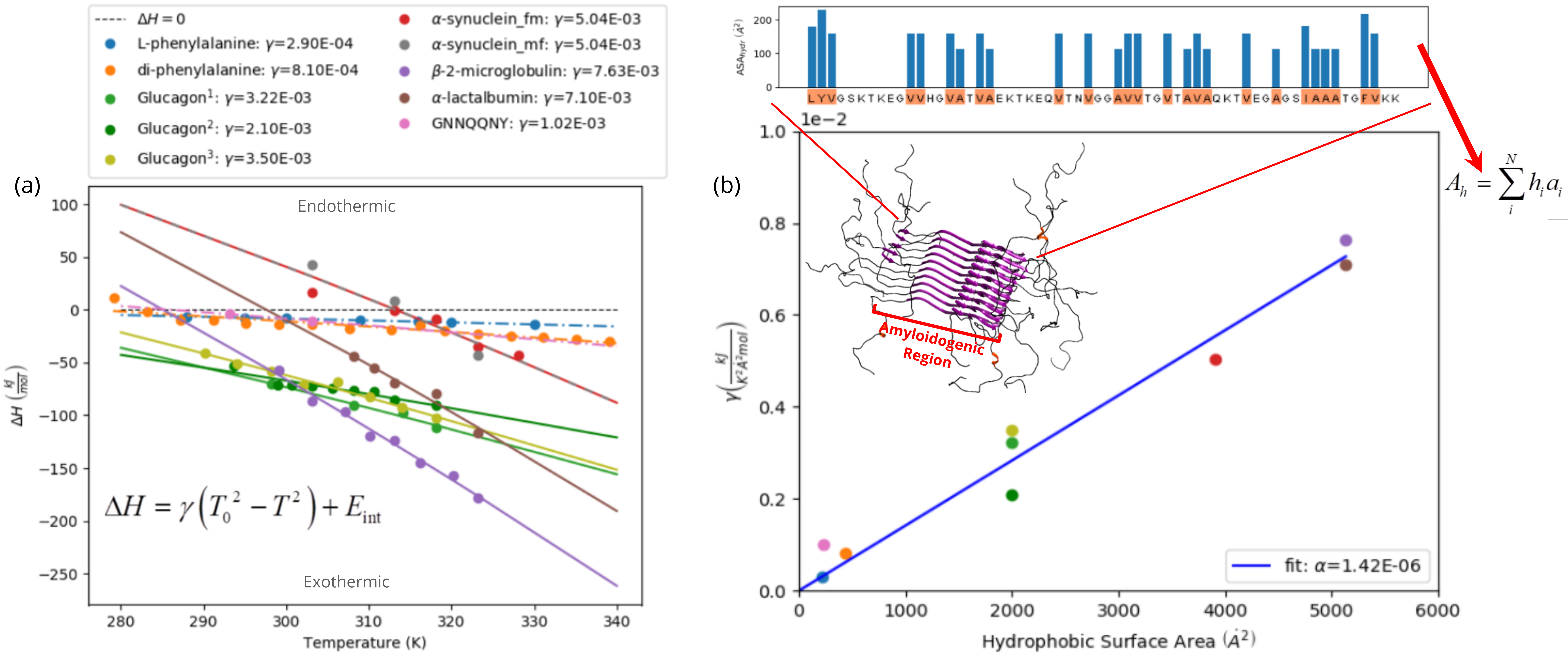}
  \caption{{\bf Enthalpy of amyloid fibril elongation and peptide and amino acid crystallisation at different temperatures}. The enthalpy of fibril elongation of $\alpha$-lactalbumin, $\alpha$-synuclein, glucagon and $\beta$-2-microglobulin~\cite{Kardos2004} (solid lines), as well as the enthalpies of crystallisation of L-phenylalanine~\cite{Kustov2013}, diphenylalanine~\cite{Mason2017} and of GNNQQNY in water (dashed lines) are shown as a function of temperature. The calorimetric measurements of glucagon$^3$ were also taken from the literature \cite{Jeppesen2010}. For $\alpha$-synuclein, ITC experiments were conducted in two set-ups: the fibrils were titrated into the monomer solution (red symbols) and the monomers were titrated into a fibril solution (grey symbols). In the case of GNNQQNY, crystals were titrated into water and their heat of dissolution was measured. Details on the experiments can be found in the SI.
  (a) Equation \ref{eq:gamma} was fitted through each of the curves to estimate the strength of the hydrophobic effect ($\gamma$) in $\frac{kJ}{K^{2} mol}$. For $T_0$ a value of 343.15K was used. (b) Relationship between the fitted values of $\gamma$ and the total hydrophobic surface area of the proteins. We calculated the total hydrophobic surface area ($A_h$) as the sum of the surface areas of all hydrophobic residues in the peptide as shown above the right panel; for more details see Supplemental Methods.
  A strong linear relationship between the hydrophobic surface area and the strength of the hydrophobic effect ($\gamma = \alpha A_h$) is shown, indicating that the contribution of the hydrophobic surface area to the strength of the hydrophobic effect ($\alpha$) is constant with a value of approximately $1.42 \cdot 10^{-6}$ $\frac{kJ}{K^2 \AA^2mol}$.
  $^1$10 mM HCl, 1 mM Na$_2$SO$_4$. 
  $^2$10 mM HCl, 1 mM Na$_2$SO$_4$, from \cite{Jeppesen2010}. 
  $^3$10 mM HCl, 30 mM NaCl.
  }
  \label{fig:ITC_data}
\end{figure*}

To probe the generality of our simulation results, we analysed isothermal titration calorimetry (ITC) measurements of the enthalpy of fibril elongation of several fibrils, both from our own experiments and from published work~\cite{Kardos2004,Jeppesen2010}, as well as calorimetric experiments of the dissolution of GNNQQNY crystals that we performed (see Methods). Furthermore, we included published calorimetric data of L-phenylalanine dissolution~\cite{Kustov2013} and a van't Hoff analysis of di-phenylalanine crystallisation~\cite{Mason2017} (Figure \ref{fig:ITC_data} a).

We find that in all cases the heat capacity of amyloid fibril elongation has negative values, see Figure \ref{fig:ITC_data} (a). In particular, also for glucagon and $\alpha$-synuclein. In the case of glucagon, we have performed the experiments under two different sets of solution conditions. Under the first condition, 10 mM HCl and 1 mM Na$_2$SO$_4$, we find that $\Delta$C$_{p,el}$ (-2.0 kJ/mol) is in excellent agreement with previous reports~\cite{Jeppesen2010} (-2.1 kJ/mol). In the second set of solution conditions (10 mM HCl and 30 mM NaCl), we find overall similar values of the heat of elongation compared to the first conditions and a slightly smaller, yet still negative, $\Delta$C$_{p,el}$ (-1.3 kJ/mol). This is surprising in the light of previous results that showed a positive value of $\Delta$C$_{p,el}$ in the presence of a higher NaCl concentration of 150 mM~\cite{Jeppesen2010}. Note that higher salt concentrations would in general be expected to reinforce the strength of the hydrophobic effect. However, at high salt concentrations and elevated temperatures, the metastability of monomeric protein solutions can be reduced, which could result in the formation of aggregates in the protein solution that is used, affecting the ITC measurements.

For $\alpha$-synuclein we also find a negative $\Delta$C$_{p,el}$ (-3.2 kJ/mol) under very similar solution conditions (PBS buffer) to where it has previously been reported to have a positive value~\cite{Ikenoue2014}. As these experiments are very sensitive to the exact state of the monomers and fibrils prior to titration, we performed the experiment in two different settings 1) by titrating fibrils into a solution of monomers and 2) by titrating a solution of monomers into seed fibrils, with consistent results, i.e. a negative $\Delta$C$_{p,el}$ (Figure \ref{fig:ITC_data} a). We also performed calorimetric experiments on GNNQQNY crystals~\cite{Nelson2005} which have so far not yet been thermodynamically characterized. We injected crystals into water and found in all cases endothermic signatures of crystal dissolution, corresponding to exothermic signatures of crystal growth (Figure \ref{fig:ITC_data} a). 

Therefore, our experimental results, together with the available data in the literature, suggest that for the large majority of experimentally accessible amyloid fibril systems, a negative value of $\Delta$C$_{p,el}$ is observed, in agreement with the predictions of our model.

\subsection*{Minimal Requirements for Cold Denaturation}
Our model shows that there are three requirements for cold denaturation to occur in amyloid systems. 

\emph{(1) The hydrophobic contribution to the fibril stability must be a substantial fraction of the net free energy of fibril formation.} 
In Figure \ref{fig:heatmap} (b) it can be observed that only versions of the model with sufficiently strong hydrophobic temperature dependence (large $\alpha$) show cold denaturation. Moreover, in our model, fibrils simulated with moderate $\alpha$ values only destabilise at temperatures around or below the freezing point (corresponding to approximately T = 0.18 in reduced units). These findings are supported by experimental measurements, showing that the elongation of the fibrils with the strongest hydrophobic core become endothermic at the highest temperatures.

\emph{(2) The overall stability of the fibril should be low to moderate.} 
This is shown in Figures \ref{fig:heatmap} (b), \ref{fig:Hbonds} and \ref{fig:NStates} (b), where only fibrils with a moderate stability show denaturation at low temperatures. Note that a fibril can have low stability due to both entropic (configurational backbone entropy) and enthalpic (hydrogen-bonding) terms, see Figures \ref{fig:Hbonds} and \ref{fig:NStates}. 

\emph{(3) The enthalpy of the fibril elongation reaction needs to change sign from being exothermic at higher temperatures to endothermic at low temperatures.} 
From Figure \ref{fig:heatmap} (b) and (c) we can observe that cold denaturation only occurs when the fibril elongation becomes endothermic at low temperatures. Note that only at temperatures below the transition from exothermic to endothermic the fibril fully destabilises. 

The free energy difference between the fibrillar state and the denatured states contains large contributions from the configurational entropy of the polypeptide chain; therefore the exact transition temperatures for cold denaturation remain difficult to estimate for any given system without detailed simulations.

Many experimental observations can be explained by these requirements. Firstly, the strict requirements can explain why only a few proteins show cold denaturation of amyloid fibrils into monomers above the freezing point \cite{Ikenoue2014}, whereas heat denaturation has been observed more generally \cite{Sasahara2005, Meersman2006, Morel2010}.
More specifically, only $\alpha$-synuclein has been reported to display full cold denaturation \cite{Ikenoue2014}. While all polypeptide systems that we have investigated display a negative $\Delta$C$_{p,el}$, we observe that the fibril growth reaction of $\alpha$-synuclein becomes endothermic at the highest temperature (Figure \ref{fig:ITC_data} a). Moreover, for $\alpha$-synuclein the transition point from an exothermic to endothermic reaction occurs around 40$^\circ$C, while the onset of cold destabilisation is around 20$^\circ$C, leading to full cold denaturation close to the freezing point~\cite{Ikenoue2014}. This is precisely as predicted by the simulations of models with a strong hydrophobic temperature dependence. In this context, it is also important to stress that amyloid fibrils of $\alpha$-synuclein have been found to be among the least stable in a large set of investigated polypeptides~\cite{Baldwin2011}, compatible with requirement two. We have performed chemical depolymerisation experiments~\cite{Narimoto2004} of glucagon and $\alpha$-lactalbumin amyloid fibrils (Figure \ref{fig:destabilisation}) and we find that both types of amyloid fibrils are significantly more stable than the ones formed by $\alpha$-synuclein.

The overall stability of amyloid fibrils corresponds to the net balance between the stabilising and destabilising factors. It has been proposed that electrostatic interactions are responsible for the cold denaturation of amyloid fibrils~\cite{Ikenoue2014}. Note that the nature of such destabilisation, over the full temperature range, does not need to originate from the hydrophobic effect, as shown by the simulations (Figures \ref{fig:Hbonds} and \ref{fig:NStates}). Destabilisation of the fibril by charges \cite{Shammas2011a} as a suggested prerequisite, but not necessarily cause, of cold denaturation of $\alpha$-synuclein~\cite{Ikenoue2014}, is therefore fully compatible with the results presented here.

\subsection*{Hydrophobicity Dominates the Thermodynamic Signature of Amyloid Fibrils}
\label{sec:value_alpha}
One can wonder to what extent the hydrophobic temperature dependence dominates the observed thermodynamic signature for fibril elongation. Therefore, the question arises as to whether other non-covalent interactions, such as charge interactions or hydrogen-bonding, may be responsible for the observation of a negative $\Delta$C$_{p,el}$~\cite{Cooper2001} in fibril elongation.

Firstly, it should be noted that mostly for small hydrophobic solutes strong negative heat capacities upon desolvation have been measured~\cite{Makhatadze1994, Kustov2013} with additional evidence for phenylalanine and di-phenylalaline presented in Figure \ref{fig:ITC_data} (a). On the other hand, electrostatic interactions in solution~\cite{Gitlin2006} and  hydrogen-bond forming substances show a weak temperature dependence for the solvation enthalpy. The temperature dependence of the hydrophobic effect is much more dramatic.

To test whether hydrophobicity can be the origin of the strong negative $\Delta$C$_{p,el}$ of fibril growth, we considered the total hydrophobic surface area per peptide buried upon assembly (amyloid fibril or crystal growth, see Methods and Supplemental Methods) and associated this directly with the model for hydrophobic temperature dependence used in our simulations. Using equation \ref{eq:gamma} we can analyse the differential enthalpy of amyloid fibril growth and peptide assembly with temperature. By fitting this equation to the experimental data of the temperature dependence of the enthalpies of assembly, we obtain an estimate of the strength of the hydrophobic effect ($\gamma$) for each of the assembling systems, see Figure \ref{fig:ITC_data} (a).

To determine the relationship between $\gamma$ and the buried hydrophobic surface area upon fibril growth, we calculated the total linear hydrophobic surface area of the peptides in the aggregating regions (see Supplemental Methods and Table \ref{tab:sequences}). As shown in the figure \ref{fig:ITC_data} (b), the relationship between $\gamma$ and the aggregating hydrophobic surface area is linear, suggesting a direct dependence of the enthalpic signature on the hydrophobic surface area. Using this fit, we obtain a strength of $ 1.42 \cdot 10^{-6} \frac{J}{\text{mol} K^2 \AA^2}$ for the hydrophobic effect per unit of hydrophobic surface area. Note that the fit through zero matches the experimental data, as is physically expected. These results strongly suggest that the general enthalpic signature of peptide assembly is directly associated with the hydrophobic temperature dependence. This conclusion is also fully compatible with previous results on the importance of the hydrophobic effect for the free energy barriers of amyloid fibril elongation~\cite{Buell2012}, as well as with recent results from atomistic simulations that show that desolvation is the main driving force for amyloid fibril formation by the A$\beta$ peptide~\cite{Schwierz2016}. The importance of the hydrophobic effect for amyloid fibril stability is also likely to explain the finding that amyloid fibrils interact with lipids and biological membranes \cite{Gellermann2005, Goodchild2014, Borro2017}.
However, it is well-known that some amino acid sequences that are not usually classified as hydrophobic, such as poly-glutamine and GNNQQNY peptides can show $\beta$-strand dominated self-assembly. We show that even for the hydrophilic GNNQQNY peptide the hydrophobic signature for $\Delta$C$_{p,el}$ holds: as shown in Figure \ref{fig:ITC_data}, the value of $\gamma$ for GNNQQNY is small, similar to that of di-phenylalanine, despite the substantial difference in molecular weight. 
As for the amyloid fibril formation by polyglutamine, it has been reported that monomeric polyglutamine peptides are considerably more compact in water compared to a fully denatured protein~\cite{Crick2006}, i.e. that water is a bad solvent for polyglutamine. Therefore, the predictions of our model might have validity even beyond the systems that are traditionally classified as hydrophobic.

\section*{Conclusion}
In this work, we use a simple model to interpret experimental observations on the temperature dependencies of the stability of amyloid fibrils and their enthalpy of elongation. Crucial to this model is the interplay between the hydrophobic temperature dependence and the flexibility of the peptide chain, the latter capturing the chain configurational entropy. This allows us to consider various enthalpy-entropy compensating mechanisms at a wide range of temperatures.  

%The model allows us to interpret the model both at low and high temperatures. 

In summary, our model shows that a large hydrophobic component of stability, low to moderate overall stability and a shift from exothermic to endothermic elongation of amyloid fibrils are necessary components to allow cold denaturation. Additionally, a strong temperature dependence of the enthalpy of fibril elongation is confirmed by ITC experiments, both from our own measurements and prior publications. Finally, the signature of the hydrophobic effect is visible in the negative $\Delta C_p$ of elongation for the large majority of investigated fibril systems. The magnitude of this negative heat capacity correlates closely with the hydrophobic surface area that is buried upon fibril formation. Hence, by delineating the necessary components for cold denaturation of amyloid fibrils, we can shed light on the crucial contribution of hydrophobicity to amyloid fibril stability and explain the observed enthalpic signature of amyloid fibril elongation.

\section*{Acknowledgements}
SA and JvG thank NWO for funding received under project number 680-91-112. AKB thanks the EMBO and Magdalene College, Cambridge for funding. HM thanks the CNRS InFinity program and the Exploratory Research Space of the RWTH Aachen University for a Theodore von K\'{a}rm\'{a}n Fellowship. We thank Alexandra Ziemski for help with the production of $\alpha$-synuclein.

\section*{Author contributions}
EvD, AKB and SA designed the research; EvD, JvG, HM, SA and AKB performed the simulations and developed the theory; AP, AH, AKB performed the experiments; EvD, AP, AH, GG, JvG, HM, DEO, AKB and SA analysed the data; and EvD, JvG, AKB and SA wrote the paper, all authors commented on the manuscript.

\section*{Competing interests}
The authors declare no competing interests.
\section*{Materials and Methods}
The code of the model and the data and scripts used to generate the figures are available on GitHub: \url{https://github.com/ibivu/amyloid_hydrophobicity}

\subsection*{The Peptide Model}
Experimentally, it has been shown that amyloid formation is usually strongly accelerated in the presence of preformed fibrillar aggregates, or seeds \cite{Buell2014}. In this study, we focus on the thermodynamic properties of elongation, or the addition of a single protein molecule to the end of a 'seed' fibril. We study this process using a cubic lattice model, where each amino acid occupies a single grid site. In this model, inspired by previous work on protein lattice models~\cite{Shakhnovich1993a, Shakhnovich1993b, Shakhnovich1994, Sali1994, Coluzza2003, Coluzza2004}, each amino acid interacts with the amino acids directly adjacent to it. If no amino acid is present, the grid site is assumed to be occupied by the solvent~\cite{Abeln2011}.

The side chain is modelled by giving each amino acid an orientation and allowing hydrogen bonds to be formed only when the side chains point in the same direction, allowing a reasonable steric approximation of $\beta$-strands~\cite{Abeln2014}.
For the peptides we used the same sequences as in ref. \cite{Abeln2014}: TFTFTFT. To investigate the effect of a lower stability, we replaced the phenylalanine residues with leucine residues, yielding the sequence TLTLTLT. The default parameters for the model are comparable to settings that are required to obtain self-replication in a simpler model~\cite{Saric2016}.

In our simulations, a seed for the study of the elongation process is represented by a pre-formed fibril consisting of 8 peptide molecules. This fibril is 'frozen' during our simulations, meaning that only moves translating or rotating the entire seed fibril are allowed. However, all interactions of the fibril with the environment are still present. Two additional monomeric protein molecules are present in the simulation box, which are allowed to make regular moves, and can attach and detach from the seed during the simulation. This setup allows us to investigate the addition of one layer (consisting of two molecules) to the pre-formed seed fibril.
In the simulations, we observe four distinct states: monomeric, amorphous, fibrillar and fully aggregated. We define these states based on the total number of external contacts (see the Supplementary Methods).

\subsection*{Temperature Dependence of the Hydrophobic Effect}
The temperature dependent solvation term $\Phi_{\text{solvent}}(T)$ accounts for the temperature dependence of the hydrophobic effect and is based on our previous work ~\cite{vanDijk2016}.

Previously, a similar solvation term allowed us to account for the temperature dependence of the hydrophobic effect in protein folding; the single and two-state models for the solvation term gave similar results for thermodynamic signatures of protein folding~\cite{vanDijk2016}. Here we use the simplest formulation of the model - with fewer parameters - that is based on a single-state representation for a side-chain with the solvent.
$\Phi_{\text{solvent}}(T)$, expressed in contributions of single residues, takes the form:
\begin{equation}
\Phi_{\text{solvent}}(T) = \sum_{i}^{N}{F_{\text{hydr}}(i) + \epsilon_{a_i,\text{solv}} K_{i,\text{solv}}  }
\label{eq:TempDepHamiltonian}
\end{equation}
\noindent Here $N$ is the total number of residues in the peptide; the term $\epsilon_{a_i,\text{solv}}$ does not depend on the temperature, whereas the free energy term, $F_{\text{hydr}}(i)$, does. $K_{i,\text{solv}}$ indicates whether or not an interaction between the solvent and residue $i$ occurs. In short, $K_{i,\text{solv}}=1$ when the side chain points in the direction of the solvent, and $K_{i,\text{solv}}=0$ otherwise. We can now incorporate the hydrophobic temperature dependence in terms of $F_{\text{hydr}}(i)$ as:
\begin{equation}
F_{\text{hydr}}(i) =  -\alpha C_h (T-T_{0})^2
\label{eq:hG}
\end{equation}
Where $T_0$ sets the temperature at which the hydrophobic effect is maximal. $\alpha$ sets the strength of the hydrophobic effect and $C_h$ indicates whether the residue makes a hydrophobic contact. $C_h$ calculated as $h_{a_i} K_{i,\text{solv}}$, where $h_{a_i}$ indicates if the amino acid is hydrophobic ($\epsilon_{a_i,w}>0$). In our potential, the amino acids that fulfil this condition are $a_i \in \{C,F,L,W,V,I,M,Y,A\}$.

\subsection*{Description of the Peptide Model}
\label{sec:math_model}
A full description of the model is given by the following equation:
\begin{equation}
\mathcal{H} = E_{\text{hb}} + E_{\text{steric}}  + E_{\text{state}} + E_{\text{aa}} + \Phi_{\text{solvent}}(T)
\label{eq:HamiltonianAggregation}
\end{equation}
\noindent This Hamiltonian, $\mathcal{H}$ is given in reduced units (k$_B$T units). The term $E_{\text{aa}}$ represents the sum of the classical pairwise amino acid interactions $\epsilon_{a_i,a_j}$, that are used in most coarse-grained simulations. The term $E_{\text{hb}}$ represents the interactions originating from hydrogen bonds between side chains of two amino acids. %Note that, since our simulation model has an implicit solvent, $\epsilon_{\text{hb}}$ indicates the difference in energy between a hydrogen bond formed in bulk water and between amino acids.
$E_{\text{state}}$ represents the energy gained from $\beta$-sheet formation. $\Phi_{\text{solvent}}(T)$ is the interaction of an amino acid with the solvent. The first four terms are kept identical to the model described by Abeln et al.~\cite{Abeln2014}. Note that only $\Phi_{solvent}(T)$ depends on the temperature, whereas the other four terms do not. Hence, we can also describe the Hamiltonian in terms of a temperature-independent and a temperature-dependent term $\mathcal{H} = E_{int} + \Phi_{solvent}(T)$.
\subsection*{Delineation of Enthalpic and Entropic Hydrophobic Contributions}
From the hydrophobic free energy contribution, as given by eqn. \ref{eq:hG}, we can also deduce the enthalpic contribution. Multiplying both sides by $\beta$($=\frac{1}{T}$), taking derivative with respect to $\beta$~ and using that $\frac{d\beta F} {d\beta} = \langle E \rangle$ we obtain:
\begin{equation}
\langle E_\text{hydr}(i) \rangle = -\alpha C_h \left( T_{0}^2 -T^2 \right)
\label{eq:hE}
\end{equation}
\noindent This equation can be used to estimate the difference in hydrophobic enthalpy between two states, based on the difference in hydrophobic residues pointing towards the solvent (see Supplementary Methods). Note that we can now also compute the entropic contribution using:
\begin{equation}
-T S_\text{hydr}(i) = F_\text{hydr}(i) - E_\text{hydr}(i)
\label{eq:hS}
\end{equation}

\subsection*{Theoretical Estimates for Hydrophobic Contributions}
\label{sec:thermodynamic_estimates}
We also try to derive theoretical estimates the hydrophobic contributions to the free energy, entropy and enthalpy of fibril elongation. These are calculated for the different types of fibrils simulated.
If we consider only the contributions that dominate the free energy differences between states in the simulations, the free energy difference of fibril elongation can be approximated by:
\begin{equation}
\Delta G = \Delta G_{\text{hydr}} + \Delta G_{\text{int}} + \Delta G_{\text{chain}}
\label{eq:SimpleFreEnergy}
\end{equation}
We can simplify this further by noting that the interaction energies, including the backbone hydrogen bonds, do not have a temperature dependence in our model. Even if we consider more broadly defined states, the enthalpy component remains approximately constant with respect to temperature for fibril elongation, if we do not take the hydrophobic temperature dependence into account. Moreover, the main contribution of the chain free energy is entropic, which turns out to be strongly temperature dependent in our simulations.
\begin{equation}
\Delta G = \Delta G_{\text{hydr}} + \Delta E_{\text{int}} - T\Delta S_{\text{chain}}
\end{equation}
In order to make a theoretical estimate, we can consider the case where the hydrophobicity dominates the change in free energy upon fibril elongation.  Then, we can calculate an estimate for the free energy change $\Delta \hat{G}_{\text{hydr}}$ as:
\begin{equation}
\Delta \hat{G}_{\text{hydr}} = \Delta G_{\text{hydr}}(T) + \Delta E_{\text{int}}
\label{eq:dG}
\end{equation}
We can then estimate the change in enthalpy as:
\begin{equation}
\Delta \hat{E}_{\text{hydr}} = \Delta E_{\text{hydr}}(T) + \Delta E_{\text{int}}
\label{eq:dE}
\end{equation}
%
%Taking the hydrophobic temperature dependence from equation \ref{eq:hE}, we obtain:
%
%\begin{equation}
%\langle \Delta \hat{E} \rangle = \gamma \left( T_0^2 - T^2 \right) + %\Delta E_{int}
%\label{eq:gamma}
%\end{equation}
%
From these estimates we can calculate the entropic contribution as:
\begin{equation}
-T \Delta \hat{S}_{\text{hydr}} = \Delta \hat{G}_{\text{hydr}}(T) - \Delta \hat{E}_{\text{hydr}}
\label{eq:dS}
\end{equation}

\subsubsection*{Estimate for hydrophobic enthalpy in the model}
By combining equations \ref{eq:hE} and \ref{eq:dE} we can obtain a relation from which we can estimate the full enthalpic contribution due to the hydrophobic effect:
\begin{equation}
\left< \hat{E}_{hydr}\left(i\right) \right> = \gamma \left(T_0^2 - T^2\right) + E_{int}
\label{eq:gamma}
\end{equation}
\noindent where $\left< \hat{E}_{hydr}\left(i\right) \right>$ is an estimate for the hydrophobic enthalpy, $\gamma$ is the strength of the hydrophobic effect, $T_0$ is the optimal temperature for hydrophobic interactions, $T$ is the temperature and $E_{int}$ is the internal energy of the protein.
For the simulation model we have $\gamma=\alpha C_{h}$.
Now by taking $\Delta C_{h}=6$ for the difference in solvent contacts between the fibrillar and monomeric state, and taking $E_{\text{int}}$ from the constant term observed in the simulations with $\alpha=0$, we can make the estimates for the hydrophobic contributions described above.

Further details on the simulation model, including descriptions of the different fibrillar states, sampling procedures and the investigation of entropically favourable $\beta$-sheets, can be found in the supplemental methods.

\subsubsection*{Estimating the strength of the hydrophobic effect in amyloid fibrils}
\label{sec:calc_alpha}
To determine the temperature dependence of the relationship between the strength of the hydrophobic effect and the hydrophobicity of a protein, we combined our peptide model with experimental data from isothermal titration calorimetry (ITC) and equilibrium experiments of amyloid fibril growth and peptide assembly into crystals. We used a least-squares fit of the temperature dependence of the enthalpy of assembly to equation
 \ref{eq:gamma}, with $T_0=343.15 K$.
Note that for a fully atomistic representation we have $ \gamma= -\alpha A_h$, where $\alpha$ is the strength of the hydrophobic effect per unit of hydrophobic surface area and $A_h$ is the total aggregating hydrophobic surface area of the protein. To calculate $A_h$, we used
\begin{equation}
  A_h = \sum_{i}^{N}{h_i a_i}
\end{equation}
where $h_i$ indicates whether or not the amino acid at position $i$ is hydrophobic and $a_i$ indicates the surface area of the amino acid at position $i$. In this case,
\begin{equation*}
   h_i = \left\{\begin{array}{lr}
   1 & \text{if } h_i \in \left\{ C, F, L, W, V, I, M, Y, A \right\} \\
   0 & \text{otherwise} \\
   \end{array}\right\}
\end{equation*}
For the calculation of the change in hydrophobic surface area upon addition to the fibril end, we only included the contributions of those sequence regions that are part of the $\beta$-sheet core of the amyloid fibril. These regions were determined using sources from \cite{Iadanza2018, Tuttle2016, Sawaya2007, Andersen2010, McKenzie1991, Frare2006, Kustov2013, Mason2017}. For a detailed description of how these core regions were determined, see the supplemental methods.

\subsection*{Preparation of (seed) fibrils for experiments}
$\alpha$-synuclein was recombinantly expressed and purified as reported previously~\cite{Buell2014}. Solutions of $\sim$200 $\mu$M $\alpha$-synuclein in phosphate buffer saline (PBS) and 0.02\% NaN$_3$ were prepared and incubated under strong stirring at temperatures between 30-37$^{\circ}$C for several days. An AFM image of the amyloid fibrils obtained in this manner are shown in Figure \ref{fig:AFM_ITC} a). Bovine $\alpha$-lactalbumin was purchased from Sigma and used without further purification. Solutions of 200-350 $\mu$M of $\alpha$-lactalbumin in 10 mM HCl and 100 mM NaCl were incubated at 37$^{\circ}$C under constant stirring for 2-3 days. Human glucagon was a kind gift from Novo Nordisk. Glucagon was studied under two different sets of solution conditions. Solutions were prepared at $\sim$700 $\mu$M in 10 mM HCl and 1 mM Na$_2$SO$_4$ and incubated under stirring at 40$^{\circ}$C for 2 days. Alternatively, solutions were prepared at $\sim$300 $\mu$M in 10 mM HCl and 30 mM NaCl and incubated under stirring at 40$^{\circ}$C for 1 day.

\subsection*{Isothermal Titration Calorimetry (ITC) Experiments of Fibril Elongation}
We performed ITC experiments to directly probe the enthalpy change associated with the addition of a protein monomer to the end of an amyloid fibril, as a function of temperature~\cite{Kardos2004, Jeppesen2010}. An ITC experiment of fibril elongation can be carried out in two distinct ways, by injecting seed fibrils into monomeric protein solutions, or by injecting monomeric protein into seed fibril suspensions. In the present study, we have explored both types of experiments. The injection of monomeric protein into seed fibrils can be carried out repeatedly, whereas a single injection of seed fibrils into a supersaturated solution of monomers leads ultimately to a complete conversion of the solution into aggregates.
The ITC experiments of fibril elongation were carried out with VP-ITC and ITC200 instruments (Malvern Instruments, UK). In the case of $\alpha$-synuclein (VP-ITC and ITC200), both types of experiments (seed fibrils into monomer and monomer into seed fibrils) were performed, whereas in the case of glucagon (VP-ITC) and bovine $\alpha$-lactalbumin (VP-ITC), monomer titrations into fibril suspensions were used. In most cases, the monomer solutions and the fibril solutions were dialysed overnight at 4$^{\circ}$C against a large volume of the same solution in order to ensure that the heat released or consumed upon titration only corresponds to the heat of reaction and not potential heats of dilution of unbalanced salt concentrations. The seed fibrils were sonicated with a probe sonicator according to protocols similar to the ones reported in~\cite{Buell2010a} in order to maximise the seeding efficiency and accelerate the reaction. A crucial point in these experiments is that the rate of heat release or consumption is sufficiently high to produce a clearly visible peak that can be integrated to yield the total amount of heat exchanged due to fibril elongation. At higher temperatures, more heat is released (due to the strongly negative heat capacity of the elongation reaction, see results) and the rate of fibril elongation is accelerated~\cite{Buell2012}, and both of these factors are beneficial for the signal-to-noise ratio of the experiment. In order to be able to perform reliable measurements at the lower end of the temperature range investigated, the only possibility to accelerate the reaction is by increasing the number of growth competent fibril ends. Figure \ref{fig:AFM_ITC} b) shows an AFM image of sonicated $\alpha$-synuclein amyloid fibrils. The ITC experiments of fibril elongation were performed at monomer concentrations between 100-350 $\mu$M ($\alpha$-lactalbumin), between 70-300 $\mu$M (glucagon) and between 50-100 $\mu$M ($\alpha$-synuclein). Figure \ref{fig:AFM_ITC} c) shows an AFM image of $\alpha$-synuclein amyloid fibrils taken out of an ITC calorimeter after an experiment.
In each case, the heat released or consumed upon an injection of monomer or fibrils, $\Delta$Q, was determined and divided by the amount of monomer that had reacted in each case. Figure \ref{fig:ITC_raw_data} shows raw data from ITC experiments of amyloid fibril elongation for the proteins $\alpha$-lactalbumin, $\alpha$-synuclein and glucagon.
%Additionally, we calculated the strength of the hydrophobic effect of fibril elongation. Using Eqn. \ref{eq:hG}, $\frac{\delta \beta F}{\delta \beta} = \langle E \rangle$ and our estimates of the free energy contributions described in the Computational Methods (see above), we obtain:
%\begin{equation}
%\label{eq:dE_est}
%    \Delta H = \gamma \left( T_0^2 - T^2 \right) + E_{int}
%\end{equation}
%\noindent Note that instead of $-\alpha C_h$ as described previously, we use $\gamma$ to represent the strength of the hydrophobic effect. For each of the fibrils, $\gamma$ was fitted to the data to obtain an estimate for the strength of the hydrophobic effect. For $T_0$ we used 343.15K.
Furthermore, we use $\Delta$E and $C_v$ when we refer to the simulations, since they are carried out at constant volume, and $\Delta$H and $C_p$ when we refer to the experiments, since they are carried out at constant pressure. As described in \cite{vanDijk2016}, these are very similar in our setup. Finally, we use $\Delta$F to refer to the free energy defined in the model, and $\Delta$G when we refer to free energy sampled by the simulations, as in \cite{vanDijk2016}.

\subsection*{Fibril Stability in the Presence of Denaturants}
We have determined the thermodynamic stabilities of amyloid fibrils using depolymerisation experiments with a chemical denaturant, as previously described ~\cite{Narimoto2004, Baldwin2011}.
For $\alpha$-lactalbumin (Figure \ref{fig:destabilisation} a), 200 $\mu$l of 377 $\mu$M fibrils was mixed with 1000 $\mu$l solution mixture of variable ratios of (1) 10 mM HCl + 100 mM NaCl and (2) 6M GndSCN + 10 mM HCl + 100 mM NaCl, in order to obtain a series of increasing denaturant concentration, ranging from 1.2 M to 5 M. The samples were incubated for 10 days at room temperature and then centrifuged for 1 h at 25$^{\circ}$C at 40 krpm. The equilibrium concentration of soluble $\alpha$-lactalbumin in the supernatant was determined using the Bradford test. The use of Bradford reagent was necessary, as the denaturant GndSCN has a considerable absorbance at 280 nm and therefore interferes with concentration determination through absorbance measurements at 280 nm, as well as with measurements of intrinsic fluorescence (see description for glucagon below). The samples were diluted 1:25 into the Bradford reagent and the values of absorbance at 595 nm were compared with a standard curve.
In the case of glucagon (Figure \ref{fig:destabilisation} b), the fibrils were depolymerised with GndHCl. 60 $\mu$l of 240 $\mu$M fibril solution (in 10 mM HCl and 30 mM NaCl) was mixed with 140 $\mu$l solution mixture of variable ratios of (1) 10 mM HCl + 30 mM NaCl and (2) 5.6 M GndHCl + 10 mM HCl + 30 mM NaCl. The samples were incubated for two days at room temperature. Then the intrinsic fluorescence was measured with a multiwell-platereader, which was equipped with an excitation filter at 280 nm, and two emission filters at 340 and 380 nm. The ratio of the fluorescence intensities at 380 nm and 340 nm was normalised for the initial and final plateau values and plotted as a function of the concentration of GndHCl. This measurement had been calibrated against the concentration of soluble peptide, determined by centrifugation and absorbance measurements.
The depolymerisation curves (Figure \ref{fig:destabilisation}) were fitted with the expression derived from the linear polymerisation model~\cite{Narimoto2004, Baldwin2011}. The free energy differences between the soluble and fibrillar states extracted from the fits are -52.5 kJ/mol for $\alpha$-lactalbumin and -58.3 kJ/mol for glucagon. These values are considerably larger in magnitude than that determined for $\alpha$-synuclein (-33.0 kJ/mol~\cite{Baldwin2011}). %Therefore, the finding that $\alpha$-synuclein displays cold denaturation can be explained by the relatively small stability of its fibrils.

\subsection*{Enthalpy of GNNQQNY crystal growth}
GNNQQNY peptide was bought from Bachem (Basel, Switzerland) as TFA salt and used without further purification. The peptide was dissolved in hot water at a concentration of 2-3 mM and left to cool down. Crystal formation occurred spontaneously after some time, but could be strongly accelerated by probe sonication. The stock solutions were kept for several days at room temperature.
A concentrated, super-critical solution of GNNQQNY peptide was found not to remain metastable for a sufficient amount of time to be used in seeded ITC experiments, such as the ones described above for amyloid fibrils. Therefore, the experiments to determine the heat of formation of GNNQQNY crystals as a function of temperature were designed as crystal dissolution experiments, rather than crystal growth experiments.  Before each calorimetry experiments, the suspensions were re-sonicated for 2 min to homogenise them. Different dilutions of the freshly sonicated stock solutions were loaded into the syringe of a VP-ITC and the cell filled with pure water. Then large injections of up to 100 $\mu$l were performed. The injection of crystals into water always yielded strongly endothermic signatures. It was noticed that identical injections yielded strongly differing peak integrals. This is likely due to the sedimentation of the crystals inside the injection syringe, which leads to the fact that injections of equal volumes do not correspond to equal quantities of injected peptide crystals.
Therefore, in order to determine the enthalpy of crystal dissolution, the total integrated heat of all injections was determined and this value was correlated with the final peptide concentration in the cell after the experiment, which was measured by UV absorption measurements at 280 nm (molar extinction coefficient 1280 M$^{-1}$ cm$^{-1}$).
When a suspension of crystals is injected into water, this corresponds to the injection of the crystals themselves, plus the injection of monomer at the critical concentration. In order to be able to determine the net heat of crystal dissolution, the obtained experimental values need to be corrected for the values of the heat of dilution of the present monomer. In order to quantify this correction, samples of crystal suspensions were heated to 60$^{\circ}$C for 30 min, which led to the dissolution of the crystals. The resulting supersaturated solution was loaded into the needle of the ITC calorimeter and the same pattern of injections was performed as for the crystals. In some cases, the re-crystallisation started inside the needle, which lead to a drift in the baseline. However, in all cases, the injection of “melted” crystals led to an exothermic signal. After these control experiments, the concentration of the peptide inside the cell was measured again.
It was found that at 20$^{\circ}$C, the heat of dilution of monomer was -187.4 $\mu$cal for a final concentration of 273 $\mu$M. The critical concentration of GNNQQNY was found to be $\sim$110 $\mu$M at 20$^{\circ}$C. Assuming that the heat of dilution scales linearly with final concentration, we subtracted -13$\mu$J from each experimental data point at 20$^{\circ}$C.
Furthermore, it was found that at 30$^{\circ}$C, the heat of dilution of monomer was -213.1 $\mu$cal for an injected concentration of 305 $\mu$M, which yielded a final concentration of 53 $\mu$M. The critical concentration of GNNQQNY was found to be $\sim$154 $\mu$M at 30$^{\circ}$. We therefore subtracted -108 $\mu$J from each experimental data point at 30$^{\circ}$C.
The measured heats of crystal dissolution are then plotted against the concentration in the ITC cell at the end of the experiment. A linear fit forced to go through 0/0 yields the molar heat of crystal dissolution. We find 4.2 cal/mol and 10.3 cal/mol at 20$^{\circ}$C and 30$^{\circ}$C, respectively, yielding a value of 0.61 cal/(mol K) or 2.6 J/(mol K) as value for the molar heat capacity of crystal dissolution. We fitted Equation \ref{eq:gamma} through the data points to estimate $\gamma$. Note that in this case, we only have data at two temperatures and therefore we cannot estimate the accuracy of the fit. However, this data and analysis does provide a good idea about the magnitude of the enthalpy and heat capacity of GNNQQNY assembly into crystals.

%\section*{References}
%\bibliography{amhydr}

%% Adds the main heading for the SI text. Comment out this line if you do not have any supporting information text.

\appendix

\renewcommand{\thefigure}{S\arabic{figure}}
\renewcommand{\theequation}{S\arabic{equation}}
\renewcommand{\thetable}{S\arabic{table}}
\setcounter{figure}{0}
\setcounter{equation}{0}
\setcounter{table}{0}

\section*{\LARGE Supporting Information}
%\section*{Heading}
%\subsection*{Subhead}
%Type or paste text here. You may break this section up into subheads as needed (e.g., one section on ``Materials'' and one on ``Methods'').

%\subsection*{Materials}
%Add a Materials subsection if you need to.

\section*{Supplemental Methods}

\subsubsection*{Definition of States}
\label{sec:states}
We observe four distinct states, which for the analysis are defined based on the number of external contacts for the two free peptides; external contacts ($C_{\text{ext}}$) are here defined as the interchain contacts between either the free peptides or the free peptides and the seed. Note that there are a further 70 external contacts present in the seed fibril that cannot be changed during the simulations.
The monomeric, fibrillar and amorphous states are then defined as:
\begin{equation}
\text{monomer} =
\begin{cases}
%1  \text{ if } \hfill C_{\text{ext}} >= 88 \\
%0 \text{ if }  C_{\text{ext}} < 88 \\
%1  \text{ if } \hfill C_{\text{ext}} = 70 \\
%0 \text{ if }  \text{otherwise} \\
1  \text{ if } \hfill C_{\text{ext}} = 0 \\
0 \text{ if }  \text{otherwise} \\
\end{cases}
\end{equation}
\begin{equation}
\text{amorphous} =
\begin{cases}
%1  \text{ if } \hfill C_{\text{ext}} >= 88 \\
%0 \text{ if }  C_{\text{ext}} < 88 \\
%1  \text{ if } \hfill  81 < C_{\text{ext}} <= 84 \\
%0 \text{ if }  \text{otherwise} \\
1  \text{ if } \hfill  11 < C_{\text{ext}} <= 14 \\
0 \text{ if }  \text{otherwise} \\
\end{cases}
\end{equation}
\begin{equation}
\text{fibril} =
\begin{cases}
%1  \text{ if } \hfill C_{\text{ext}} >= 88 \\
%0 \text{ if }  C_{\text{ext}} < 88 \\
%1  \text{ if } \hfill C_{\text{ext}} >= 86 \\
%0 \text{ if }  \text{otherwise}\\
1  \text{ if } \hfill C_{\text{ext}} >= 16 \\
0 \text{ if }  \text{otherwise}\\
\end{cases}
\end{equation}
\begin{equation}
\text{fully aggregated} =
\begin{cases}
%1  \text{ if } \hfill C_{\text{ext}} >= 88 \\
%0 \text{ if }  C_{\text{ext}} < 88 \\
%1  \text{ if } \hfill  81 < C_{\text{ext}} <= 84 \\
%0 \text{ if }  \text{otherwise} \\
1  \text{ if } \hfill C_{\text{ext}} = 21 \\
0 \text{ if }  \text{otherwise} \\
\end{cases}
\end{equation}

\subsection*{Enthalpy Sampling}
For the simulation we calculate the enthalpy difference between the ensemble of the fully aggregated state ($A: C_{\text{ext}}=21$) and the monomeric state ($M: C_{\text{ext}}=0$); in the latter case only the seed peptides still make contacts. The enthalpy difference $\Delta E$ is then calculated as $\Delta E  = \langle E \rangle_{A} - \langle E\rangle_{M}$.

\subsection*{Hydrogen Bonds}
The term describing hydrogen bonds can be written as $E_{\text{hb}}= \sum \epsilon_{hb} H_{i,j} \cdot C_{i,j}$, where $\epsilon_{\text{hb}}$ represents the potential energy per hydrogen bond. $H_{i,j}$ = 1 indicates that a hydrogen bond is present, and $H_{i,j}$ = 0 indicates that no hydrogen bond is present. In our model, the water interactions are implicit, so $\epsilon_{hb}$ indicates the difference between a hydrogen bond of an amino acid with the solvent and a hydrogen bond in bulk water. Hydrogen bonds between amino acids and the solvent are typically stronger than the hydrogen bonds in a bulk solvent. We investigate the cases $\epsilon_{\text{hb}} \in \{0.25,0.5,0.75,1.0\}$ with $\epsilon_{\text{hb}}=0.5$ as default.

\subsection*{Entropically Favourable $\beta$-sheets}
\label{sec:S_beta}
We also investigated the effect of an entropic bonus for $\beta$-sheets. The high stability of $\beta$-sheets at elevated temperatures suggests that $\beta$-sheets may have a higher entropy than other secondary structure elements. Moreover, amino acids with a high propensity for $\beta$ strand formation tend to be $\beta$-branched \cite{Chou1983}, suggesting that the $\beta$-strand states will be entropically more favourable. Core regions of amyloid forming proteins tend to have a high $\beta$-strand forming propensity~\cite{Trovato2006a}, to model this propensity we allow an entropic term, $N_{\beta}$, to be set for the $\beta$-strand state modelling degeneracy of the state.
This allows us to investigate the effect of a local entropic `bonus' a residue receives for being in a $\beta$-sheet. Unless otherwise stated $N_{\beta}=1$, giving no bias.

\subsection*{Sampling Analysis}
\label{sec:umbrella}
We used the umbrella sampling method to sample the conformational space~\cite{Grossfield2003}. As order parameter, we used the number of external contacts, $C_{\text{ext}}$. We use a quadratic biasing potential to define $E_{\text{umbr}}$:
\begin{equation}
E_{\text{umbr}} = \mathcal{H} + k (C_{\text{ext}}-C_{\text{ext},0})^2
\end{equation}
\noindent Where $k$ is the spring constant, $\mathcal{H}$ the Hamiltonian defined in eqn. \eqref{eq:HamiltonianAggregation} (see Methods), and $C_{\text{ext},0}$ the value towards which the simulation is biased. In our simulations, $k=2$ and $C_{\text{ext},0} \in \{0,5,10,15,20,25\}$.

\subsubsection*{Identification of the core regions of the amyloid fibrils} 
To calculate the change in accessible hydrophobic surface area upon amyloid fibril elongation, we determined the fibril core-forming regions of each protein. PDB structures of the aggregating fibrils were available $\beta$-2-microglobulin (PDB ID:6GK3 \cite{Iadanza2018}), $\alpha$-synuclein (PDB ID:2N0A \cite{Tuttle2016}, residues 38-97) and for the microcrystals formed by the GNNQQNY peptide from Sup35 (PDB ID:2OMM \cite{Sawaya2007}). In the case of glucagon, it has been reported that the fibril is formed by nearly the entire sequence of 29 amino acids~\cite{Andersen2010}. For bovine $\alpha$-lactalbumin, no high resolution fibril structure is available to-date. However, $\alpha$-lactalbumin is homologous to lysozyme~\cite{McKenzie1991}, for which the aggregating region is known from limited proteolysis experiments~\cite{Frare2006}. Therefore, we queried both protein sequences using PSI-BLAST (accessed 25 January 2019) on the Swissprot database with default settings for two iterations. We selected the first twelve hits or the $\alpha$-lactalbumin query (Uniprot: P00709, Q9N2G9, P00712, P00711, P09462, Q9TSN6, Q9TSR4, P00714, P61633, P30201, P61626, P00716) and added the lysozyme sequence and two hits of the lysozyme query with 70-80 percent sequence identity with lysozyme (Uniprot: P04421, P61631, P79811) to ensure that both proteins were represented in the sequence set. Subsequently, we performed multiple sequence alignment on these sequences using Clustal Omega with default settings. The resulting alignment was used to predict the aggregating region of $\alpha$-lactalbumin from the known aggregating region of lysozyme. The estimated aggregating region was TFHT…GINY of $\alpha$-lactalbumin (Uniprot P00711, region 48-122).
%These data were taken from the literature for $\alpha$-synuclein~\cite{Vilar2008} and $\beta$-2-microglobulin~\cite{Monti2005}. For $\alpha$-lactalbumin, the core residues of the fibrils have not yet been mapped. As an approximation, we use the same core residues as have been determined for lysozyme~\cite{Frare2006}, given the close relationship between the two proteins~\cite{McKenzie1991}.
In order to cover a wider range of total hydrophobic surface area, we also included thermodynamic data of the self-assembly of a hydrophobic amino acid \cite[phenylalanine,][]{Kustov2013} and of a hydrophobic dipeptide \cite[diphenylalanine,][]{Mason2017}. The temperature dependent enthalpy of dissolution of solid phenylalanine, the reverse of the enthalpy of assembly, was measured calorimetrically~\cite{Kustov2013}, whereas for diphenylalanine, a van't Hoff analysis of the solubility was performed~\cite{Mason2017}. In both cases, the entire hydrophobic surface area was assumed to become buried upon assembly, given that both components assemble into crystals, rather than amyloid fibrils.

\section*{Supplementary data}

\subsection*{Simulations with model parameters affecting the enthalpic and entropic contributions to fibril stability}
We also investigated how the cold denaturation and destabilisation depend on the different model parameters. In our model we can adjust the stability of the fibril by changing the strength of the hydrogen bonds (Figure \ref{fig:Hbonds}). The stability obtained through H-bonds is purely enthalpic in our model. Similarly, we can change the entropic contribution to fibril stability by changing the $\beta$-strand propensity (Figure \ref{fig:NStates}); this term is purely entropic in our model. The simulations show that the stability of aggregates at low temperatures depends on the stability at physiological temperatures, see Figure  \ref{fig:Hbonds} and \ref{fig:NStates} (b).

\subsection*{Additional experimental data}
Figure \ref{fig:ITC_raw_data} shows the raw data from some of the ITC experiments used to produce the plots in Figure 5 of the main manuscript. The details about the injection volumes and monomer concentrations can be found in the figure caption. In order to analyse these data, a baseline was defined and each of the peaks due to the injection of monomer was integrated numerically. In addition, we have also acquired atomic force microscopy images (Figure \ref{fig:AFM_ITC}), which illustrated both the preparation of the seed fibrils of $\alpha$-synuclein through sonication, which increases the seeding efficiency~\cite{Buell2014}, as well as the subsequent growth during the ITC experiment upon injection of soluble protein.

\subsection*{Fibril stability in the presence of denaturants}
Our model predicts that cold denaturation of amyloid fibrils is only observed for fibrillar systems that are not very highly thermodynamically stable with respect to the soluble state of the protein. This condition arises from the requirement that the low temperature-induced weakening of the hydrophobic effect leads to a sufficient overall loss in stability for the equilibrium concentration of soluble protein to increase substantially. 

Published data show that $\alpha$-synuclein amyloid fibrils are among the least stable amyloid fibrils characterised to-date, in agreement with the observation that these fibrils are among the few that have been found to display cold denaturation. We have performed additional experiments on the thermodynamic stabilities of glucagon and $\alpha$-lactalbumin amyloid fibrils under the same conditions under which we have performed the ITC experiments (Figure \ref{fig:destabilisation}).

\clearpage
\newpage

\begin{figure*}
\centering
\includegraphics[width=0.6\linewidth]{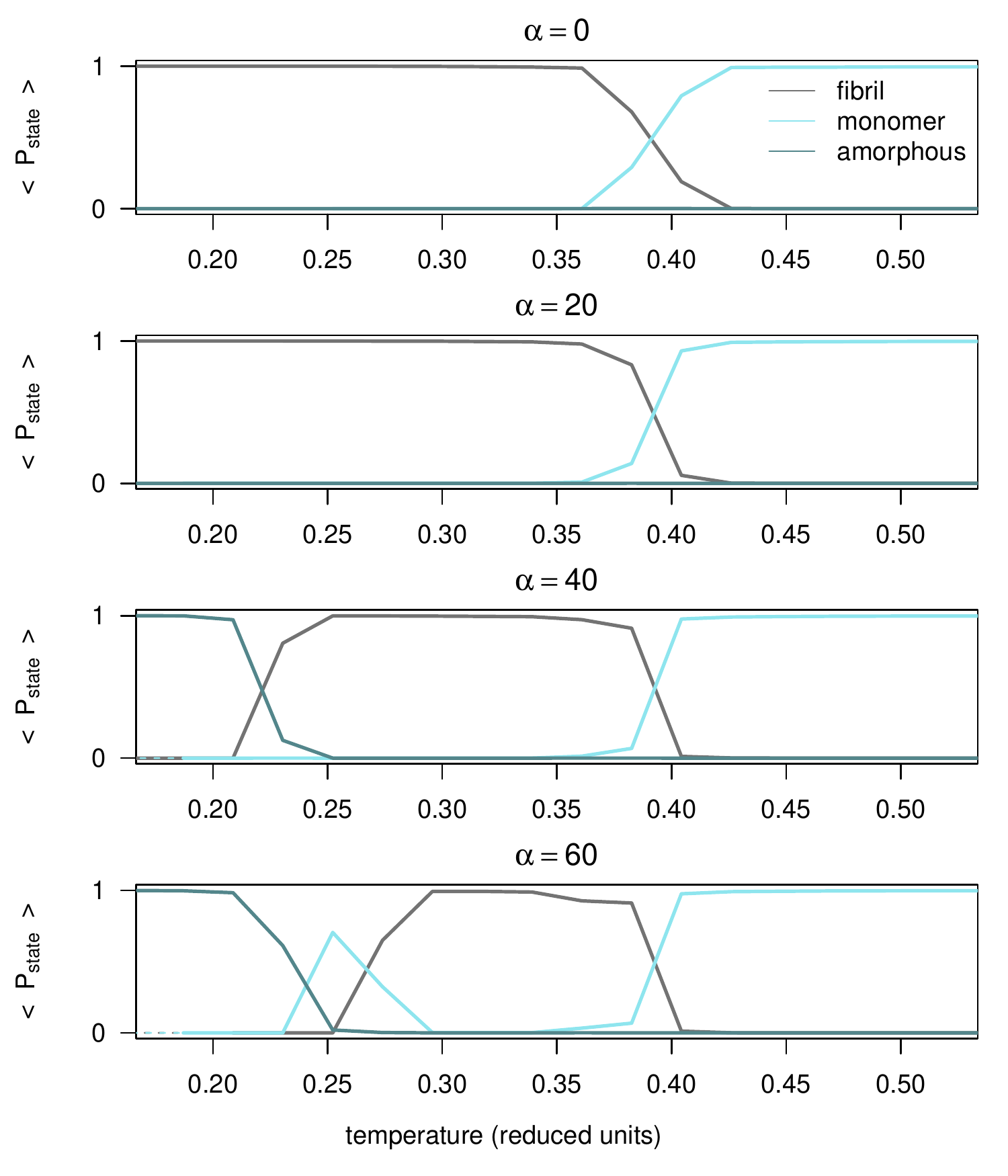}
\caption{{\bf State diagrams for fibril elongation.}
Here, we explored the effect of the strength of the hydrophobic temperature dependence, $\alpha$, on the stability of the aggregates. Three different states can be discerned: the fully formed fibril (black), denaturation of the fibril into monomers (cyan) and an amorphous aggregate where the two additional layers are not fully formed. Only in the models with a hydrophobic temperature dependence, cold destabilisation ($\alpha$ = 40, 60), or cold denaturation into monomers, may be observed ($\alpha$ = 60). The dashed lines indicate that the state has not been observed (sampled) in the simulations at the corresponding temperature. Note that the reduced temperature units for this model can be interpreted to have a freezing point around T = 0.18, and boiling point just above T = 0.4.
}
\label{fig:characteristicsPlot}
\end{figure*}

\clearpage
\newpage

\begin{figure*}[t]
\centering
\includegraphics[width=0.5\linewidth]{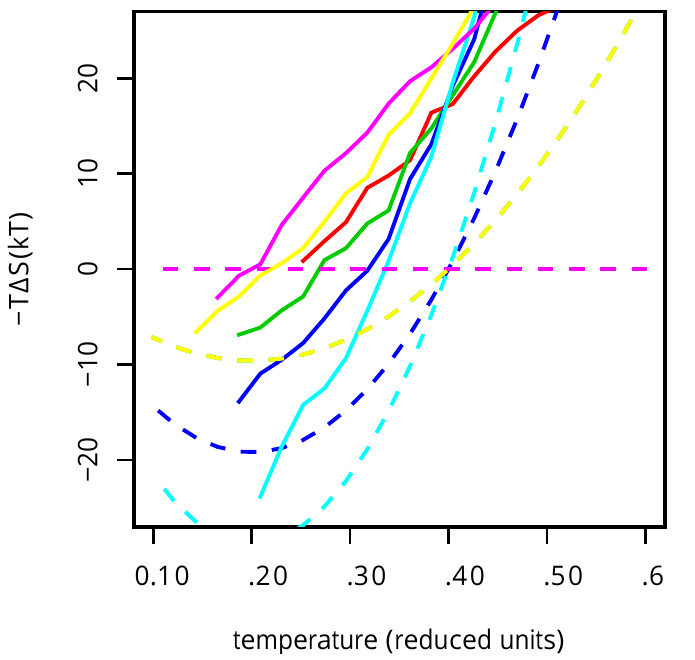}
\caption{Entropy difference between the monomeric state and the fully aggregated state as a function of temperature for different strengths of the hydrophobic effect, as in \ref{fig:heatmap} (b) and (c). One can see that the entropy differences predicted to be caused by the hydrophobic effect (dotted lines) are not able to explain the full entropy differences observed in the simulations. The estimates for the entropic contributions due to the hydrophobic effect are estimated using equation \ref{eq:dS}}.
\label{fig:TdS}
\end{figure*}

\clearpage
\newpage

\begin{figure*}[t]
\includegraphics[width=\linewidth]{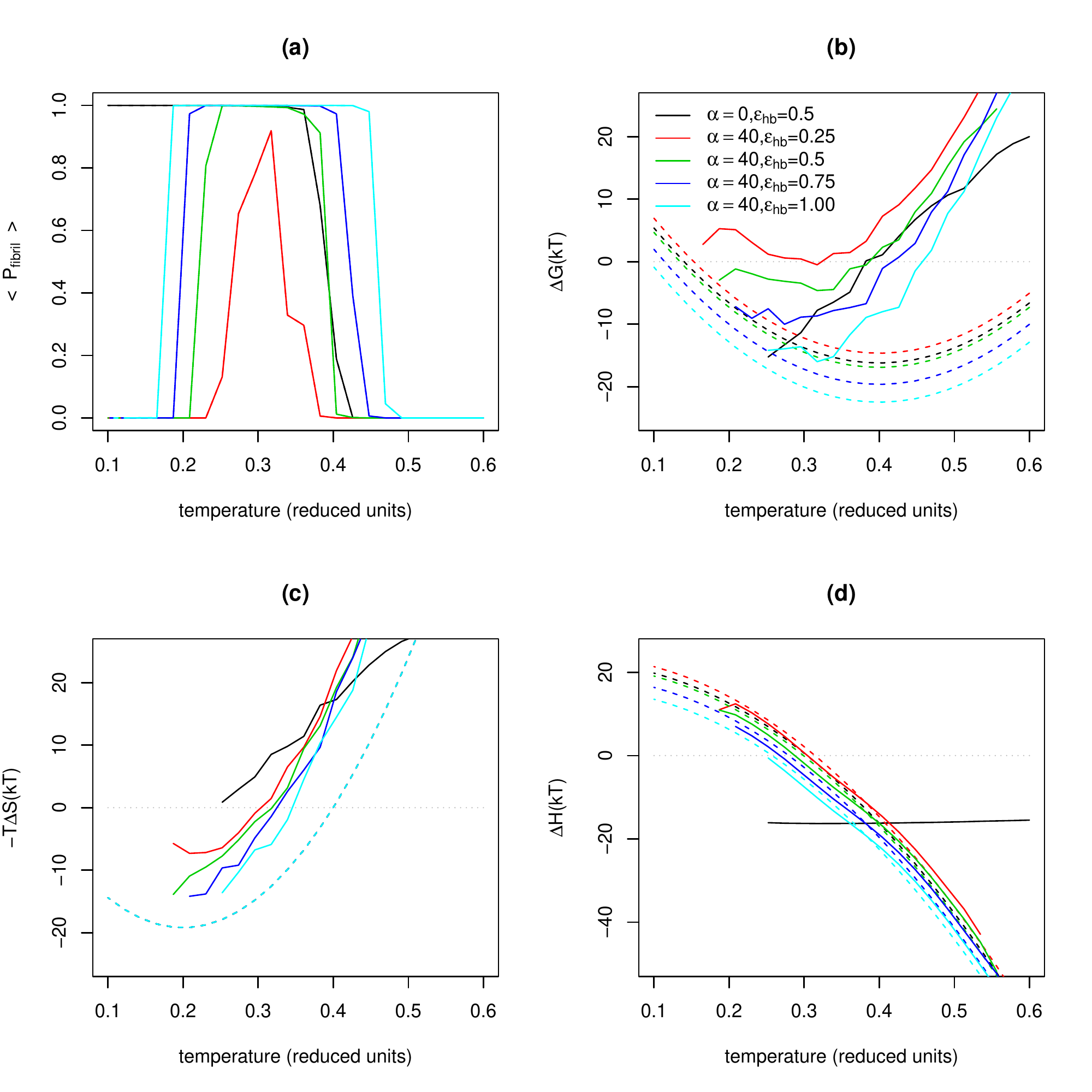}
\caption{{\bf Varying the enthalpic contribution to fibril stability through hydrogen bonds}. We explored the stability of the fibrillar state for different values of the hydrogen bond strength ($\epsilon_{hb}$) in the model. For varying values $\epsilon_{hb}$, and $\alpha=40$ the state diagram for the fibrillar state (a), the free energy (b), and corresponding entropic (c) and enthalpic (d) contributions are shown. Increasing the hydrogen bond strength makes the fibril more stable (b), resulting in a wider temperature range over which the fibrillar state is stable (a).Dotted lines indicate estimates for the hydrophobic contributions  showing  $\Delta \hat{G}_{\text{hydr}}$, $-T\Delta \hat{S}_{\text{hydr}}$ and $\Delta \hat{E}_{\text{hydr}}$; these estimates are generated using Eqns. 13, 15 and 14 with corresponding $\alpha$, $\Delta C_h=-6$ and with an offset, $E_\text{int}=\Delta H $ based on simulations with the equivalent peptide for $\alpha=0$.}
\label{fig:Hbonds}
\end{figure*}

\clearpage
\newpage

\begin{figure*}[t]
\includegraphics[width=\linewidth]{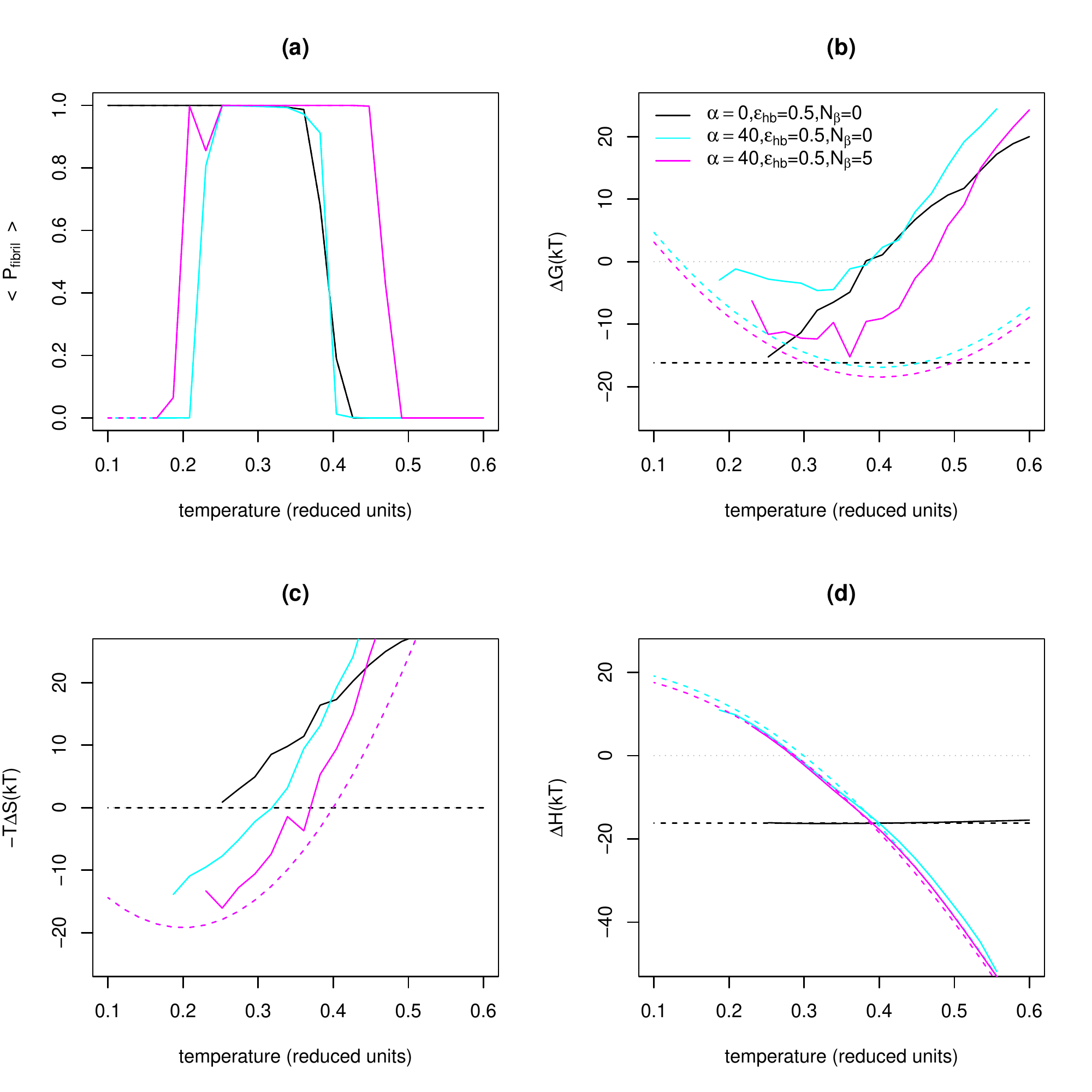}
\caption{{\bf Varying the entropic contribution to fibril stability through chain entropy}. We explored the stability of the fibrillar state for different values of the (entropic) propensity of $\beta$-strand state  ($N_\beta$) in the model. For varying values $N_\beta$, and $\alpha=40$ the state diagram for the fibrillar state (a), the free energy (b), and corresponding entropic (c) and enthalpic (d) contributions are shown. Increasing the $\beta$-strand propensity makes the fibril more stable (b), resulting in a wider temperature range over which the fibrillar state is stable (a). Dotted lines indicate estimates for the hydrophobic contributions  showing  $\Delta \hat{G}_{\text{hydr}}$, $-T\Delta \hat{S}_{\text{hydr}}$ and $\Delta \hat{E}_{\text{hydr}}$; these estimates are generated using Eqns. 13, 15 and 14 with corresponding $\alpha$, $\Delta C_h=-6$ and with an offset, $E_\text{int}=\Delta H $ based on simulations with the equivalent peptide for $\alpha=0$.}
\label{fig:NStates}
\end{figure*}

\clearpage
\newpage

\begin{figure*}[t]
\includegraphics[width=\linewidth]{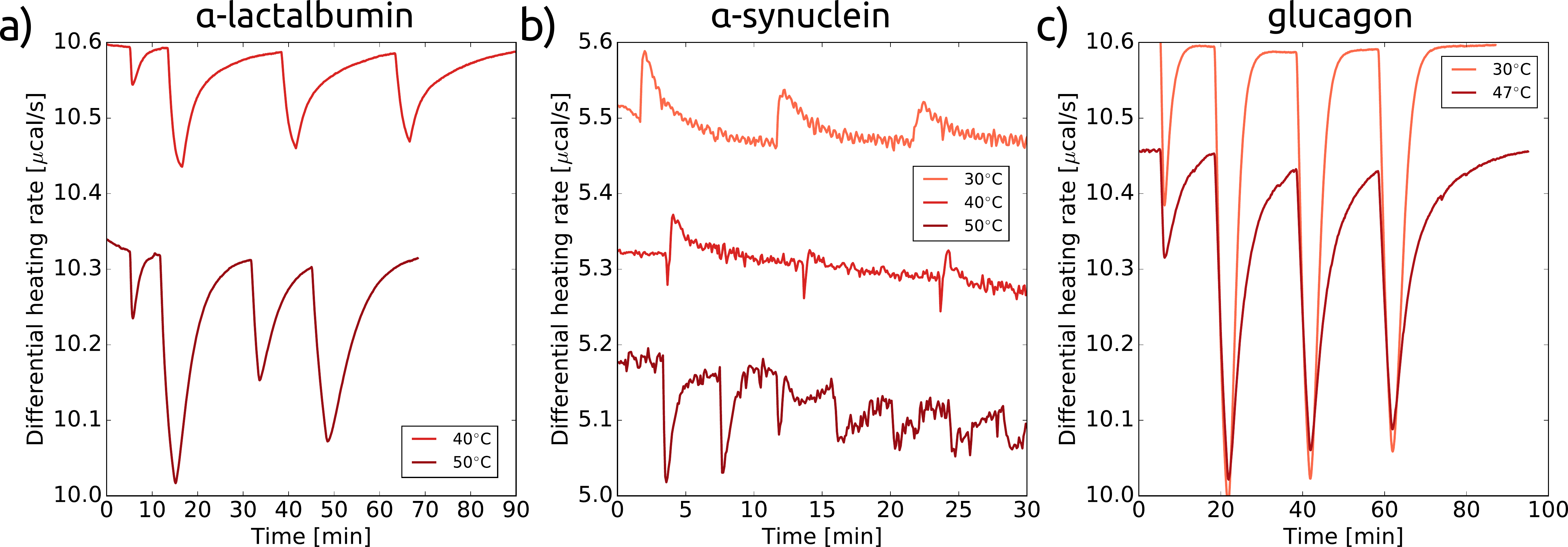}
\caption{{\bf Representative ITC raw data} Raw data of ITC experiments are shown for experiments where monomer solutions were titrated into seed fibril suspensions. Experiments were performed with a VP-ITC (a,c) and an ITC200 (c) instruments. a) Injections of 10, 80, 80, 80$\mu$l (40$^{\circ}$) and 10, 80, 40, 80$\mu$l (50$^{\circ}$) of a solution of $\alpha$-lactalbumin (50 $\mu$M in 10 mM HCl+100mM NaCl) into a suspension of sonicated seed fibrils. b) Injections of 2 $\mu$l of solutions of monomeric $\alpha$-synuclein at 380 $\mu$M (50$^{\circ}$C), 390 $\mu$M (30$^{\circ}$C) and 430 $\mu$M (40$^{\circ}$C) into seed fibril suspensions.  c) Injections of 20, 80, 80, 80$\mu$l (30 and 47$^{\circ}$) of a solution of glucagon (100 $\mu$M in 10 mM HCl+30mM NaCl) into a suspension of sonicated seed fibrils.}
\label{fig:ITC_raw_data}
\end{figure*}

\clearpage
\newpage

\begin{figure*}[t]
\includegraphics[width=\linewidth]{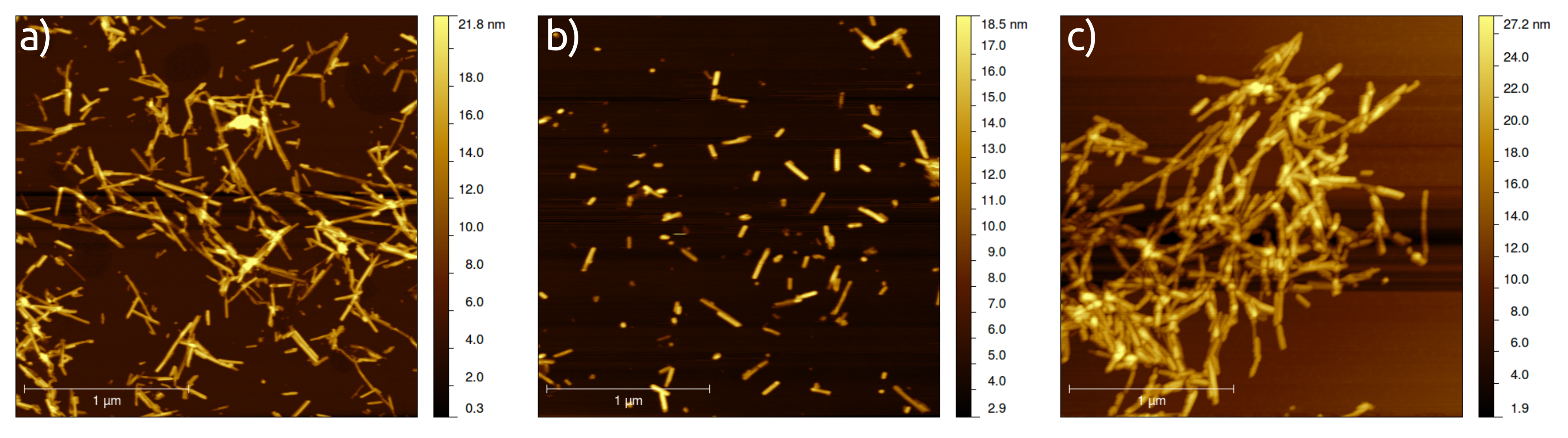}
\caption{{\bf AFM images illustrating $\alpha$-synuclein amyloid fibril elongation} Atomic force microscopy (AFM) images were taken of seed fibrils before sonication to shorten the length distribution and enhance the seeding efficiency (a), after 10 s of sonication with a sonication probe (b) and after an ITC experiment (c), where the fibrils (40 $\mu$M) had been incubated with a total of 60 $\mu$M of monomeric $\alpha$-synuclein.}
\label{fig:AFM_ITC}
\end{figure*}

\clearpage
\newpage

\begin{figure*}[t]
\centering
\includegraphics[width=0.8\linewidth]{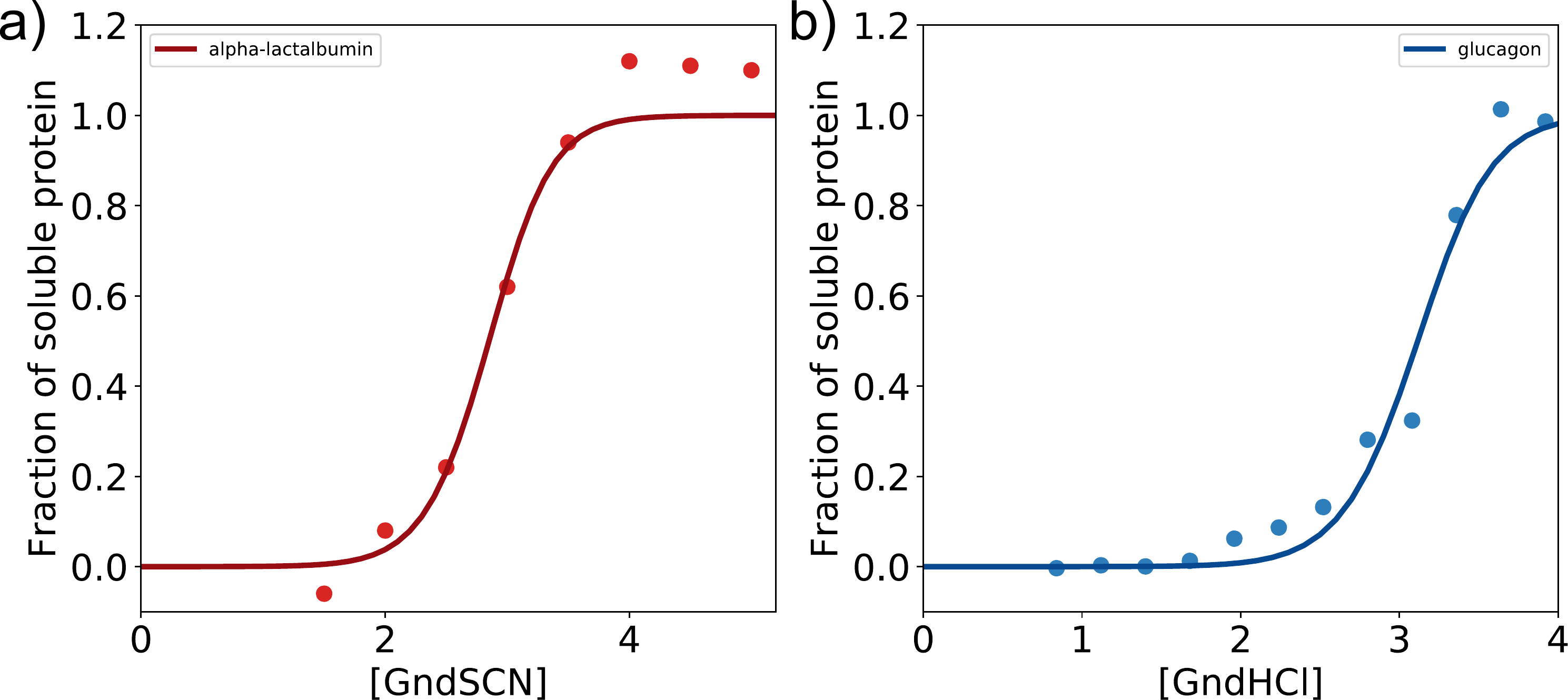}
\caption{{\bf Chemical depolymerisation of amyloid fibrils} a) $\alpha$-lactalbumin amyloid fibrils depolymerised with the strong denaturant GndSCN. b) glucagon amyloid fibrils depolymerised with GndHCl. The values of the free energy difference between the soluble and fibrillar states are -52.5 kJ/mol ($\alpha$-lactalbumin) and -58.3 kJ/mol (glucagon). These values should be compared with the one determined for the considerably less stable $\alpha$-synuclein amyloid fibrils of -33.0 kJ/mol~\cite{Baldwin2011}.}
\label{fig:destabilisation}
\end{figure*}

\clearpage
\newpage

\begin{figure*}[t]
\centering
\includegraphics[width=0.8\linewidth]{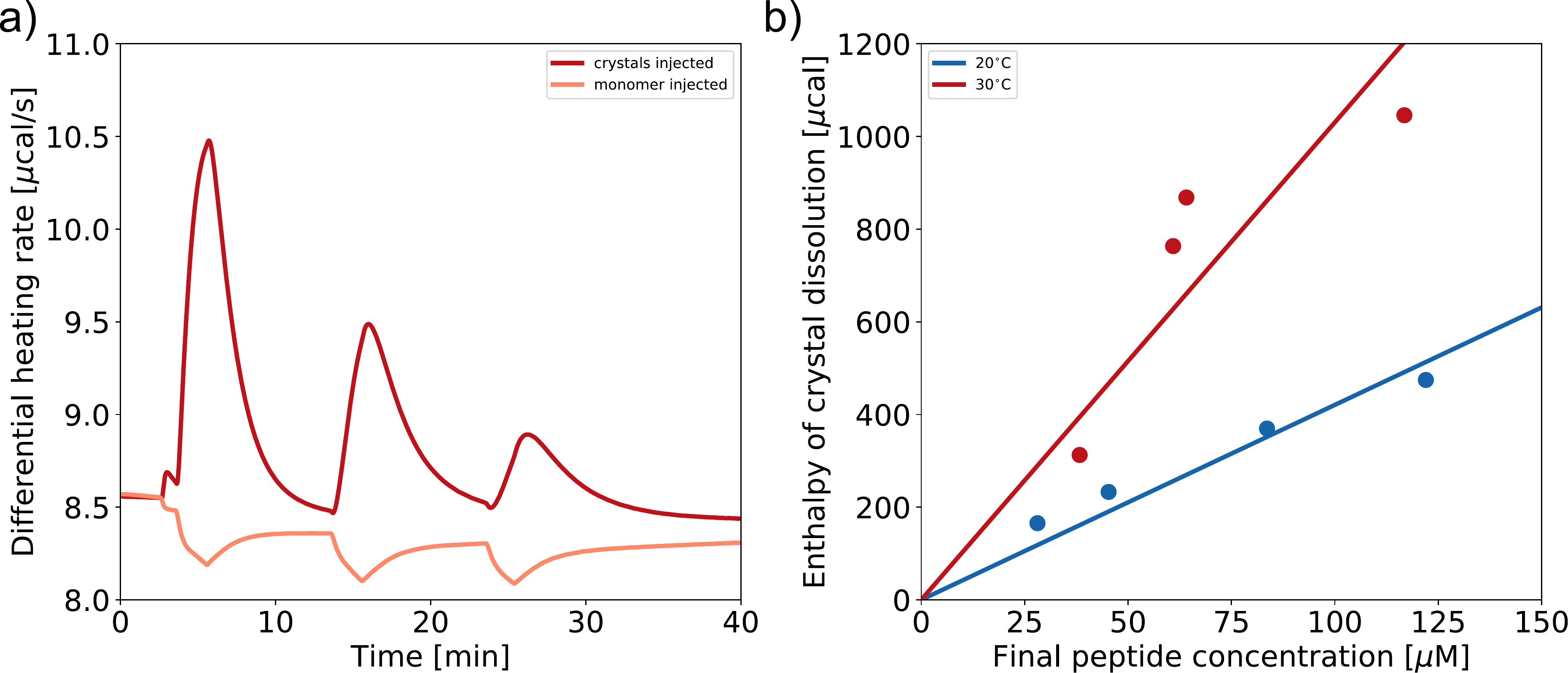}
\caption{{\bf Thermodynamics of GNNQQNY crystallisation} a) Raw ITC data of the injectionof GNNQQNY crystals and monomer into pure water. Experimental details see Methods section. b) Summary of the calorimetric results of GNNQQNY crystal dissolution. The data points are corrected for the exothermic heats of dilution of the monomeric content of each injection. The linear fits to the data sets at the two temperatures are used to determine the molar enthalpies of crystal dissolution, which corresponds to the negative of the molar heats of crystal growth.}
\label{fig:GNNQQNY}
\end{figure*}

\clearpage
\newpage

\begin{table*}
  \begin{tabular}{lllll}
   \textbf{Protein} & \textbf{$A_h$} & \textbf{$\gamma \left[ \frac{kJ}{K^{2} mol} \right]$} & \textbf{$\alpha \left[ \frac{kJ}{K^2\AA^2 mol} \right]$} & \textbf{$E_{int} \left[ kJ \right]$}\\
   GNNQQNY & 229 & $1.02\cdot10^{-3}$ & $4.45\cdot10^{-6}$ & -31.232082 \\
   L-phenylalanine & 218 & $2.90\cdot10^{-4}$ & $1.33\cdot10^{-6}$ & -16.418050 \\
   di-phenylalanine & 436 & $8.10\cdot10^{-4}$ & $1.86\cdot10^{-6}$ & -33.447338 \\
   Glucagon$^1$ & 1990 & $3.22\cdot10^{-3}$ & $1.62\cdot10^{-6}$ & -162.668964 \\
   Glucagon$^2$ & 1990 & $2.10\cdot10^{-3}$ & $1.06\cdot10^{-6}$ & -125.478512 \\
   Glucagon$^3$ & 1990 & $3.50\cdot10^{-3}$ & $1.76\cdot10^{-6}$ & -159.018135 \\
   $\alpha$-synuclein & 3906 & $5.04\cdot10^{-3}$ & $1.29\cdot10^{-6}$ & -98.911013 \\
   $\beta$-2-microglobulin & 5131 & $7.63\cdot10^{-3}$ & $1.48\cdot10^{-6}$ & -277.822212 \\
   $\alpha$-lactalbumin & 5133 & $7.10\cdot10^{-3}$ & $1.38\cdot10^{-6}$ & -205.792291
  \end{tabular}
\caption{Estimates of $\alpha$ from ITC experiments on four different proteins. Data from the experiments was fitted to \ref{eq:gamma} to estimate the value of $\alpha$.}
\label{tab:alpha}
\end{table*}
%\end{widetext}

\clearpage
\newpage

\begin{table*}
  \begin{tabular}{ll}
   \textbf{Protein} & \textbf{Sequence} \\
   L-phenylalanine & F \\
   di-phenylalanine & FF \\
   Sup35 & GNNQQNY \\
   Glucagon$^1$ & HSQGTFTSDYSKYLDSRRAQDFVQWLMNT \\
   Glucagon$^2$ & HSQGTFTSDYSKYLDSRRAQDFVQWLMNT \\
   Glucagon$^3$ & HSQGTFTSDYSKYLDSRRAQDFVQWLMNT \\
   $\alpha$-synuclein & LYVGSKTKEGVVHGVATVAEKTKEQVTNVGGAVVTGVTAVAQKTVEGAGSIAAATGFVKK \\
   $\beta$-2-microglobulin & FLNCYVSGFHPSDIEVDLLKNGERIEKVEHSDLSFSKDWSFYLLYYTEFTPTEKDEYACRVNHV \\
   $\alpha$-lactalbumin & TFHTSGYDTQAIVQNNDSTEYGLFQINNKIWCKDDQNPHSSNICNISCDKFLDDDLTDDIMCVKKILDKVGINY
  \end{tabular}
\caption{Sequences used to calculate the total hydrophobic surface area. Only the amyloigenic regions of the fibrils were used (which does not equal the full sequence for $\alpha$-lactalbumin, $\alpha$-synuclein and $\beta$-2-microglobulin).}
\label{tab:sequences}
\end{table*}
%\clearpage
%\newpage

%\bibliography{pnas-sample}
\end{document}